\documentclass[iop,letter]{emulateapj}

\newcommand{\kms}{\,km~s$^{-1}$}    \newcommand{\sqcm}{\,cm$^{-2}$}  
\newcommand{\caw}{\ion{Ca}{2}}     \newcommand{\nao}{\ion{Na}{1}}
       \newcommand{\cf}{\ion{C}{4}}
\newcommand{\sif}{\ion{Si}{4}}     \newcommand{\nf}{\ion{N}{5}}       
\newcommand{\sit}{\ion{Si}{3}}     \newcommand{\siw}{\ion{Si}{2}} 
     \newcommand{\few}{\ion{Fe}{2}} 
\newcommand{\hi}{\ion{H}{1}}       \newcommand{\hw}{\ion{H}{2}}
\newcommand{\oi}{\ion{O}{1}}       \newcommand{\tiw}{\ion{Ti}{2}}
\newcommand{\no}{\ion{N}{1}}       \newcommand{\alw}{\ion{Al}{2}}
\newcommand{\nit}{\ion{Ni}{2}}     \newcommand{\cw}{\ion{C}{2}}
\newcommand{\co}{\ion{C}{1}}       
\newcommand{\sw}{\ion{S}{2}}       \newcommand{\niw}{\ion{Ni}{2}}

\newcommand{\lya}{Lyman-$\alpha$}  
        \newcommand{\hst}{\emph{HST}}
    
\newcommand{\tm}{\tablenotemark}   \newcommand{\tn}{\tablenotetext}

\newcommand{\rbs}{\object{RBS\,144}} 
\newcommand{\fn}{\object{Fairall\,9}} 
\newcommand{\he}{\object{HE\,0056--3622}} 
\newcommand{\ngc}{\object{NGC\,7714}} 
\newcommand{\phl}{\object{PHL\,2525}} 

\begin{document}
\shorttitle{Magellanic Stream Abundances}
\shortauthors{Fox et al.}
\title{The COS/UVES Absorption Survey of the Magellanic Stream: I. 
One-Tenth Solar Abundances along the Body of the Stream}\footnotemark[1]
\footnotetext[1]{Based on observations taken under program 12604 of
  the NASA/ESA Hubble Space Telescope, obtained at the Space 
  Telescope Science Institute, which is operated by the Association of 
  Universities for Research in Astronomy, Inc., under NASA contract 
  NAS 5-26555, and under proposal ID
  085.C-0172(A) with the Ultraviolet and Visual Echelle 
  Spectrograph (UVES) on the Very Large Telescope (VLT) Unit 2
  (Kueyen) operated by the European Southern Observatory (ESO) at
  Paranal, Chile.}
\author{Andrew J. Fox$^{1}$, Philipp Richter$^{2,3}$, Bart P. Wakker$^4$, 
  Nicolas Lehner$^5$, J. Christopher Howk$^5$, Nadya Ben Bekhti$^6$,   
  Joss Bland-Hawthorn$^7$, \& Stephen Lucas$^8$}
\affil{
$^1$ Space Telescope Science Institute, 3700 San Martin Drive,
  Baltimore, MD 21218\\
$^2$ Institut f\"ur Physik und Astronomie, Universit\"at Potsdam, Haus
   28, Karl-Liebknecht-Str. 24/25, 14476, Potsdam, Germany\\
$^3$ Leibniz-Institut f\"ur Astrophysik Potsdam (AIP),
   An der Sternwarte 16, 14482 Potsdam, Germany\\
$^4$ Department of Astronomy, University of
  Wisconsin--Madison, 475 North Charter St., Madison, WI 53706\\
$^5$ Department of Physics, University of Notre Dame, 225 Nieuwland 
Science Hall, Notre Dame, IN 46556\\
$^6$ Argelander-Institut f\"ur Astronomie, Universit\"at Bonn, 
Auf dem H\"ugel 71, 53121 Bonn, Germany\\
$^7$ Institute of Astronomy, School of Physics, University of Sydney, 
NSW 2006, Australia\\
$^8$ Department of Physics \& Astronomy, University College London, 
Gower Street, London, WC1E 6BT, UK}
\email{afox@stsci.edu}

\begin{abstract}
The Magellanic Stream (MS) is a massive and extended tail of multi-phase gas 
stripped out of the Magellanic Clouds and interacting with the Galactic halo. 
In this first paper of an ongoing program to study the Stream in absorption,
we present a chemical abundance analysis based on \hst/COS and 
VLT/UVES spectra of four AGN (RBS\,144, NGC\,7714, PHL\,2525, and 
HE\,0056--3622) lying behind the MS. Two of these sightlines yield 
good MS metallicity measurements:
toward RBS\,144 we measure a low MS metallicity of 
[S/H]=[\sw/\hi]=$-$1.13$\pm$0.16 
while toward NGC\,7714 we measure
[O/H]=[\oi/\hi]=$-$1.24$\pm$0.20. 
Taken together with the published MS metallicity toward NGC\,7469, 
these measurements indicate a uniform abundance of $\approx$0.1\,solar 
along the main body of the Stream. 
This provides strong support to a scenario in which most of the Stream was
tidally stripped from the SMC $\approx$1.5--2.5\,Gyr ago
(a time at which the SMC had a metallicity of $\approx$0.1\,solar),
as predicted by several N-body simulations.
However, in Paper II of this series (Richter et al. 2013), we report a much 
higher metallicity (S/H=0.5\,solar) in the inner Stream toward Fairall\,9, 
a direction sampling a filament of the MS that Nidever et al. (2008)
claim can be traced kinematically to the LMC, not the SMC.
This shows that the bifurcation of the Stream is
evident in its metal enrichment.
Finally we measure a similar low metallicity 
[O/H]=[\oi/\hi]=$-$1.03$\pm$0.18 
in the $v_{\rm LSR}$=150\kms\ cloud toward HE\,0056--3622,
which belongs to a population of anomalous velocity clouds near the South 
Galactic Pole. This suggests these clouds are associated with the Stream or 
more distant structures (possibly the Sculptor Group, which lies in this 
direction at the same velocity), rather than tracing foreground Galactic 
material.

\end{abstract}

\section{Introduction} 
Satellite interactions are one of the important mechanisms by which 
massive galaxies acquire gas and sustain their ongoing star formation.
These interactions generate extended tidal gas streams
which pass through the galaxies' multi-phase halos toward their disks.
The gas brought in by tidal streams,
and other processes including cold and hot accretion, 
clearly plays an important role in galaxy evolution \citep[see][]{Pu12}, 
but the relative importance of these processes,
and the detailed manner in which gas enters galaxies, is poorly known.
The extended halo of the Milky Way, which contains a well-mapped population
of high-velocity clouds (HVCs) and is pierced by hundreds of quasar sightlines,
represents an ideal locale to investigate these galaxy feeding processes.

The Magellanic Stream (MS) stands as the most striking example 
of a satellite interaction in the Galactic neighborhood. 
Massive \citep[1.5--5$\times$10$^8\,M_\odot$ in \hi, depending on its 
distance;][]{Br05, Ni10, Be12}, extended 
\citep[$\approx$140\degr\ long, or $\approx$200\degr\ when including 
the Leading Arm][]{HW88, BT04, Br05, Ni10}, and of low-metallicity 
\citep[$\approx$0.1--0.5 solar;][this paper]{Lu94, Gi00, Fo10},
the MS represents a prime opportunity to study the properties,
origin, and fate of a nearby gaseous stream. 
Discovered in 21\,cm emission \citep{WW72, Ma74, Ma77}, 
the Stream exhibits a multi-phase structure
including cool molecular cores \citep{Ri01}, cold neutral  
\citep{Ma09} and warm neutral \citep{Br05} gas,
and regions of warm-ionized \citep{Lu94, Lu98, WW96, Pu03b} and
highly-ionized \citep{Se03, Fo05a, Fo10} plasma.

Substantial theoretical and simulational work has gone into understanding
the Stream's origin and fate. The favored origin mechanisms are tidal stripping
\citep{FS76, MF80, LL77, LL82, Li95, Ga94, GN96, YN03, Co06, Be10, Be12, 
DB11a, DB12}
and ram-pressure stripping \citep{Me85, MD94, HR94, Ma05, DB11b},
with perhaps some contribution from feedback from LMC and SMC stars
\citep{Ol04, Ni08}.
Following new measurements of the proper motion of the Magellanic 
Clouds \citep{Ka06a, Ka06b, Ka13, Pi08}, which are compatible
with the idea that the Magellanic Clouds are on 
their first passage around the Galaxy, the most recent tidal models
\citep{Be10, Be12, DB11a, DB12} confirm that the Stream is produced largely
by the effect of tides raised by the LMC on the SMC during a recent encounter,
as was realized earlier by \citet{GN96}.
The fate of the Stream has been investigated by \citet{BH07}; in 
their model, the MS is breaking down in a shock cascade and will subsequently 
rain onto the Galaxy in a warm ionized phase, or evaporate into the 
hot corona \citep[see also][]{HP09}.
Empirical determinations of the Stream's gas-phase properties are critical to 
test these models and thereby confirm the fate of the Stream.

With UV absorption-line studies of background quasars, a wide range of 
diagnostics of the physical and chemical conditions in the MS are available.
Early \hst\ measurements of the MS metallicity 
using the \sw/\hi\ and \siw/\hi\ ratios found values 
$Z_{\rm MS}$=0.28$\pm$0.04 solar in the \fn\ direction \citep{Gi00}, lying 
close to the LMC and SMC, and $Z_{\rm MS}$=0.25$\pm$0.07 solar in the 
NGC\,3783 direction through the Leading Arm \citep{Lu94, Lu98, Se01, Wa02}.
More recently, \citet[][hereafter F10]{Fo10} reported a lower metallicity 
of $Z_{\rm MS}$=0.10$\pm$0.03 solar in the tip of the Stream using the 
\oi/\hi\ ratio, which is robust against ionization and dust corrections.
These measurements support the view that the Stream originates 
in the SMC, which has a current-day mean oxygen abundance of 0.22 solar,
and not the LMC, which has a current-day mean oxygen abundance of 
0.46 solar \citep{RD92}, but until now, good statistics on the 
Stream's metal abundance have been lacking. 
In addition, caution must be used in comparing gas-phase abundances in the 
Stream with current-day abundances in the Clouds, 
since the Stream was stripped some time ago ($\approx$2\,Gyr in the tidal 
models), and its metallicity will reflect the metallicity of the outer
regions of its parent galaxy at the time at which it was stripped, not at the 
present day.

We have undertaken a multi-wavelength program to study the MS using optical 
and UV absorption. As part of this program, we are measuring a larger set of 
elemental abundances in the MS for comparison with SMC and LMC abundances, 
thereby providing more stringent constraints on the origins of the Stream.
In this paper, we present the results from UV and optical spectra 
of four AGN: \rbs, \ngc, \phl, and \he.
These sightlines all lie behind the Stream, as can be seen in Figure 1,
which shows a 21\,cm emission map of the entire Magellanic region generated
from the Leiden-Argentine-Bonn (LAB) survey \citep{Ka05}.
The sightlines sample a wide range of \hi\ column density, from 
log\,$N$(\hi)$_{\rm MS}$=20.17 toward \rbs\ down to 18.24 toward \phl. 
Basic properties of these sightlines
are given in Table 1. In a companion paper, \citet[][hereafter Paper II]{Ri13}, 
we present detailed results from the \fn\ sightline, lying only 8.3\degr\ away 
on the sky from \rbs.

\begin{deluxetable*}{lcccc cc c ccc cc}
\tabletypesize{\footnotesize}
\tablecaption{Sightline Properties and Details of Observations}
\tablehead{Target & Type & $l$ & $b$ & $\alpha_{\rm SMC}^{\rm a}$ 
& $v_0$(\hi)$_{\rm MS}^{\rm b}$ & \multicolumn{2}{c}{\underline{~~~log\,$N$(\hi)$_{\rm MS}^{\rm c}$~~~}} 
& \multicolumn{3}{c}{\underline{~~~~~~Exposure Time (s)~~~~~~}}
& \multicolumn{2}{c}{\underline{~~~~~~~S/N$^{\rm f}$~~~~~~~}}\\
& & (\degr) & (\degr) & (\degr) & (km\,s$^{-1}$) & GASS & LAB & UVES\tm{d} & G130M\tm{e} & G160M\tm{e} & 1300\,\AA\ & 1550\,\AA}
\startdata
\rbs\ & Sey1 & 299.485    & $-$65.836 & 21.6 &$\phn$$\phn$92 & 20.17   & 20.27 & 6000 & 2352 & 2972 & 34 & 24\\ 
\ngc\ & Nuc\tm{g}  &$\phn$88.216& $-$55.564 & 75.9 & --320         & 19.09   & 18.93 & \nodata\tm{h}& 2062 & 2760 & 20 & 15\\
\phl\ & QSO  &$\phn$80.683& $-$71.317 & 60.5 & --260         & $<$18.21 (3$\sigma$)& 18.24 & 5605 & 2146 & 2772 & 22 & 19\\
\he\  & Sey1 & 293.719    & $-$80.898 & 36.7 &$\phn$150\tm{i}& 18.70\tm{i}& 18.87\tm{i} & 5400 & 4598 & 5657 & 24 & 22
\enddata
\vspace{-1.5cm}
\tn{a}{Angular separation of sightline from center of SMC.}\\
\tn{b}{Central LSR velocity of \hi\ emission from the MS (or, in the case of \he, from the AVC).}\\
\tn{c}{Logarithmic \hi\ column density in the Stream measured from GASS (14.4\arcmin\ beam) and LAB (30\arcmin\ beam) surveys.}\\
\tn{d}{VLT/UVES exposure time with Dichroic 1 390+580 setting.}\\ 
\tn{e}{\hst/COS exposure time with G130M/1289 and G160/1589 settings.}\\  
\tn{f}{Signal-to-noise ratio per resolution element of COS data at 1300\,\AA\ and 1550\,\AA.}
\tn{g}{Extended galaxy nucleus, not point-source.} 
\tn{h}{No UVES data taken for this target.}
\tn{i}{The absorption centered at 150\kms\ toward \he\ traces the AVCs near the South Galactic Pole, not the MS.}
\end{deluxetable*}

\begin{figure}
\epsscale{0.9}
\plotone{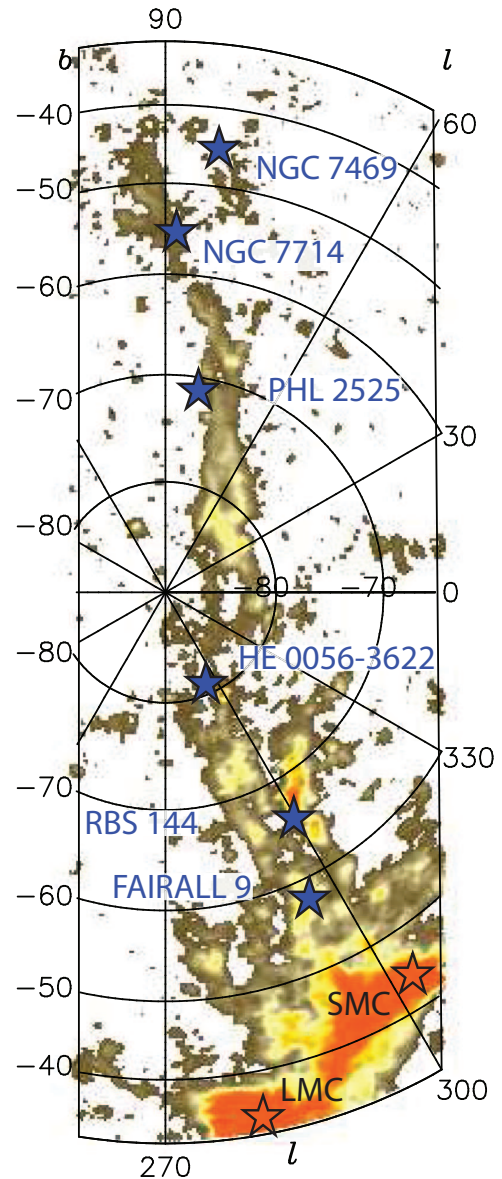} 
\caption{\hi\ 21\,cm map of the MS generated from the LAB survey
and color-coded by \hi\ column density. 
The color scale ranges from 5$\times$10$^{18}$ to 3$\times$10$^{20}$\sqcm.
The map is shown in Galactic coordinates centered on the South Galactic Pole.
The integration range in deviation velocity is $v_{\rm dev}$=$-$500 to $-$80
and 50 to 500\kms, chosen to minimize contamination by foreground emission.
The four directions from this paper plus Fairall\,9 (Paper II) and 
NGC\,7469 (F10) are marked. Note how the RBS\,144 and Fairall\,9 directions 
sample the two principal filaments of the Stream.}
\end{figure}

We use the term ``main body'' of the Stream to refer to its principal 
\hi-emitting filaments passing from the LMC and SMC through the 
South Galactic Pole and over into the Western Galactic hemisphere 
(see Figure 1). The term does not refer to the Leading Arm or 
Interface Region of the Magellanic System, which are discussed in \citet{Br05}.
21\,cm studies have shown the main body to be bifurcated both spatially 
\citep{Pu03a} and kinematically \citep[][hereafter N08]{Ni08} into 
two principal filaments.  
N08 reported that one of the two filaments (hereafter the LMC filament)
can be traced kinematically back to a star-forming area 
of the LMC known as the South-Eastern \hi\ Overdensity (SEHO) region,
which contains the giant \hw\ region \object{30\,Doradus}, but that
the other filament (hereafter the second filament) 
cannot be traced to either Magellanic Cloud, so its origin
until now has been unknown.

We note that \rbs\ lies behind the second filament reported by N08, 
whereas \phl\ and \ngc\ lie near the tip of the Stream, where the filamentary
structure is difficult to discern.
We also note that toward \he, the Stream is centered
near $v_{\rm LSR}$=--10\kms, where MS absorption overlaps with foreground
Galactic absorption. For this reason \he\ is not a good probe of the MS,
even though the sightline passes through it.
However, the \he\ spectrum shows high-velocity absorption at LSR velocities of
80--200\kms\ (see \S3.5), similar to the velocities of the 
``Anomalous velocity clouds'' (AVCs) found 
in \hi\ emission in directions (like \he) near the 
South Galactic Pole, where the Sculptor Group of galaxies is located
\citep{Ma75, HR79, Pu03a}.
\he\ is therefore a useful probe of the AVCs. 
We use metallicity measurements to investigate the origin of these clouds, 
including the possibility that they are associated with the MS.

This paper is structured as follows. In \S2 we describe the observations
and data reduction. In \S3 we describe the measurement techniques,
and briefly discuss the observed properties of absorption in the four
directions under study. In \S4, we derive and discuss the chemical 
abundances in the MS, analyze the dust content of the Stream,
and present \emph{Cloudy} models to investigate the effects
of ionization. A discussion on the implications of these
results on the origin of the Stream is presented in \S5. We summarize 
our findings in \S6. Throughout this paper, we present velocities in the 
kinematical local-standard-of-rest (LSR) reference frame, we take atomic 
data from \citet{Mo03}, and we follow the standard definition of abundances
[X/H]$\equiv$log\,(X/H)--log\,(X/H)$_{\odot}$. We adopt solar
(photospheric) abundances from \citet{As09}. 

\section{Observations}
\hst/COS observations of the four target AGN were taken
under \hst\ program ID 12604 (PI: A. Fox), using the G130M/1291 and G160M/1589 
settings, together giving wavelength coverage from 1132--1760\,\AA.
Details of these observations are given in Table 1.
The COS instrument is described in \citet{Gr12}.
In all cases, the different exposures were taken at different FP-POS positions,
to move the spectra on the detector and mitigate the effects of fixed-pattern 
noise. 
The CALCOS pipeline (v2.17.3) was used to process and combine the raw data,
yielding a set of co-added {\tt x1dsum.fits} files that were used 
for the analysis. No velocity offsets larger than a few \kms\ were found 
between interstellar absorption features in these spectra, i.e. 
there was no evidence 
for significant velocity-scale calibration errors. The COS spectra have a 
velocity resolution $R\approx$16\,000 (instrumental FWHM$\approx$19\kms), and 
were rebinned by three pixels to $\approx$7.0\kms\ pixels for display
and measurement.
The S/N ratios of each spectra measured at 1300\,\AA\ 
(G130M grating) and at 1550\,\AA\ (G160M grating) are given in Table 1.
A second, orbital night-only reduction of the data was completed, in which the 
data were extracted only over those time intervals when the Sun altitude 
(FITS keyword {\tt SUN\_ALT}) was less than 20\degr. This reduction is 
necessary to reduce the strong geocoronal emission (airglow) in 
\oi\ 1302, \siw\ 1304, and \lya\ at velocities of $|v_{\rm LSR}|\!\la\!200$\kms.
Finally, the spectra were normalized around each line of interest using 
low-order polynomial fits to the local continuum. 

Three of the four target AGN (all except \ngc) were observed with VLT/UVES
under ESO program ID 085.C-0172(A) (PI: A. Fox). The UVES instrument is 
described in \citet{De00}. These observations were taken in Service Mode
with Dichroic 1 in the 390+580 setting, no binning, a 0.6\arcsec\
slit, and under good seeing conditions (FWHM $<$0.8\arcsec), giving 
wavelength coverage from 3260-4450\,\AA\ and 4760--6840\,\AA.
The spectral resolution in this setting ($R\!\approx\!70\,000$) corresponds to 
a velocity resolution FWHM=4.3\kms. The data were reduced using the standard 
UVES pipeline \citep[based on][]{Ba00} running in the Common Pipeline Library 
(CPL) environment, using calibration frames taken close in time to the 
corresponding science frames.
The S/N per resolution element of the UVES data at 3930\,\AA\ (near \caw)
is $\approx$68 toward \rbs, $\approx$40 toward \he, and $\approx$101 
toward \phl.

For our 21\,cm analysis, we use spectra from two radio surveys, the 
Leiden-Argentine-Bonn (LAB) survey \citep{Ka05} and the GASS survey
\citep{MG09, Ka10}\footnotemark[2]\footnotetext[2]{Data from both surveys are 
available at http://www.astro.uni-bonn.de/hisurvey/profile.},
using the closest pointings to our target directions.
Using measurements from two radio telescopes allows us to investigate beamsize 
effects, which limit the precision with which one can derive the \hi\ 
column density (and, in turn, the metallicity) in a pencil-beam direction. 
The LAB survey observations were taken with the 30\,m Villa Elisa telescope 
with a beam size of 30\arcmin. The GASS observations were taken with the 
64\,m Parkes telescope with a beam size of 14.4\arcmin.

The \hi\ columns in the Stream were calculated from the equation 
$N$(\hi)=$1.823\times10^{18}\,{\rm cm}^{-2} \int^{v_{\rm max}}_{v_{\rm min}} T_b\,{\rm d}v$ 
\citep[e.g.][]{DL90}, where $T_b$ is the brightness temperature in Kelvin, 
$v_{\rm min}$ and $v_{\rm max}$ are the velocity integration limits
in \kms, and the line is assumed to be optically thin.
The differences in the \hi\ columns derived from the LAB and GASS spectra 
are visible in Table 1. We adopt these differences as the
beamsize uncertainty in the \hi\ column density in the MS
along the line-of-sight to each AGN.
This uncertainty translates directly to the derived metal 
abundances, and we include it as a systematic error in our results, although 
toward \rbs\ we are able to confirm the MS \hi\ column 
in the pencil-beam line-of-sight
using a fit to the damping wings of \lya\ (see \S3.2).

\section{Magellanic Stream Absorption}
Figure 2 shows the VLT/UVES data covering
\caw, \nao, and \tiw\ for the \rbs, \phl, and \he\ directions.
Figures 3, 4, 5, and 6 show the \hst/COS data for the same three directions
plus \ngc, for which no UVES data are available.
In each case we include the 21\,cm emission-line profile for comparison,
from either the GASS or LAB survey.
The 21\,cm data are shown unbinned in Figure 2, but for display are 
rebinned to five pixels in Figures 3 to 6.
We now discuss the measurement techniques and present an overview of the 
MS (or AVC) absorption seen in each direction.

\begin{figure*}
\epsscale{1.25}
\plotone{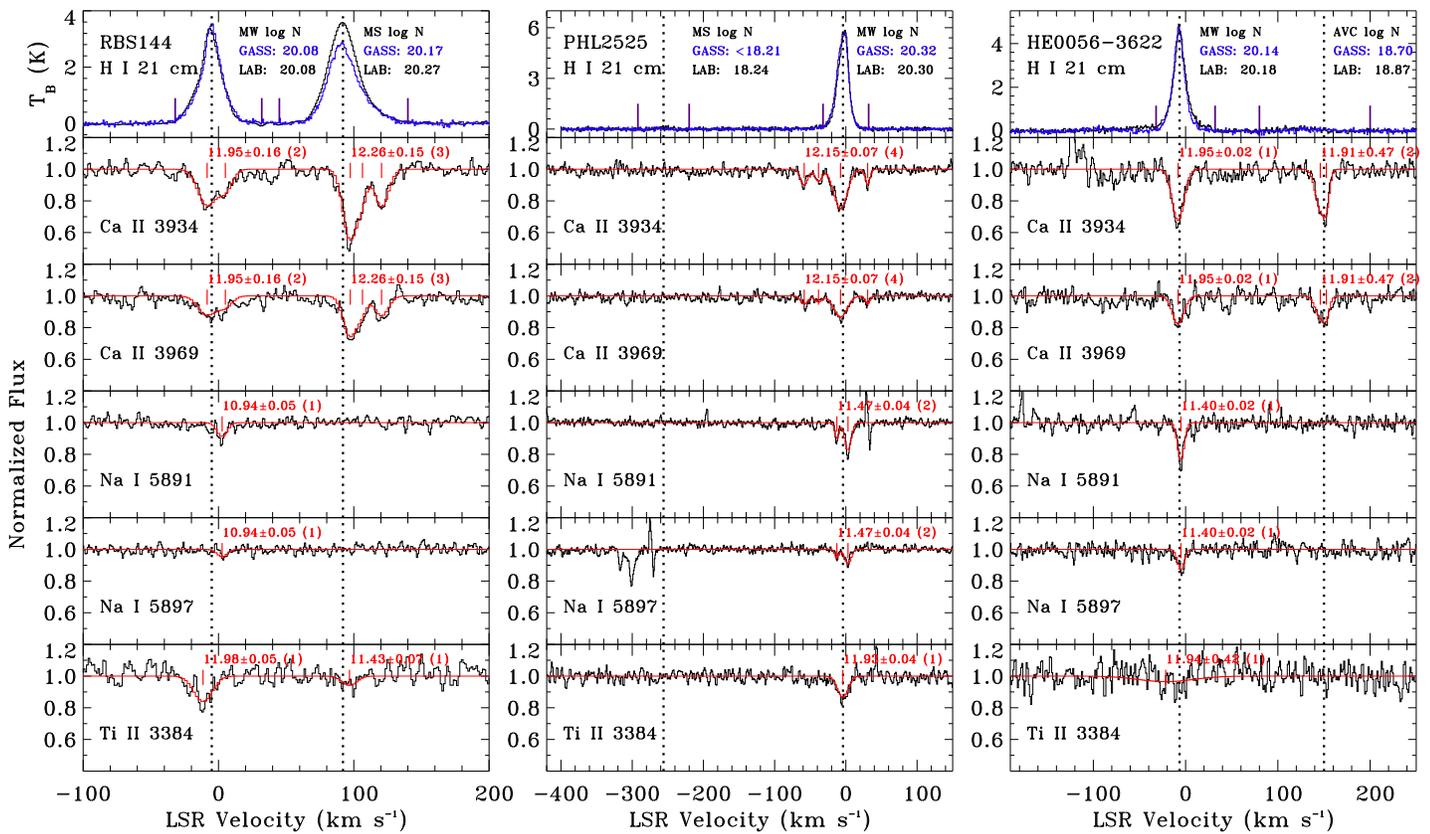} 
\caption{VLT/UVES spectra of RBS\,144, PHL\,2525, and HE\,0556--3622.
 The top panels show the 21\,cm \hi\ spectra 
 from the LAB (black) and GASS (blue) surveys. 
 All other panels show the UVES spectra in black, and our  
 Voigt-profile fits convolved with the line spread function in red. 
 The total column densities in the MW, MS, and/or AVC components are 
 annotated above the spectra, with 
 the number of velocity components given in parentheses. 
 Red tick marks show the centroids of each velocity component.
 We fit the \tiw\ MS absorption toward RBS\,144 with a single 
 component (bottom-left panel), but it is not a 3$\sigma$ detection, 
 so we treat is as an upper limit in the analysis. 
 The vertical dotted lines show the velocity centroids of the MW and MS
 21 cm emission components. Note the compressed y-scale on each panel.}
\end{figure*}

\begin{figure*}
\epsscale{1.15}
\plotone{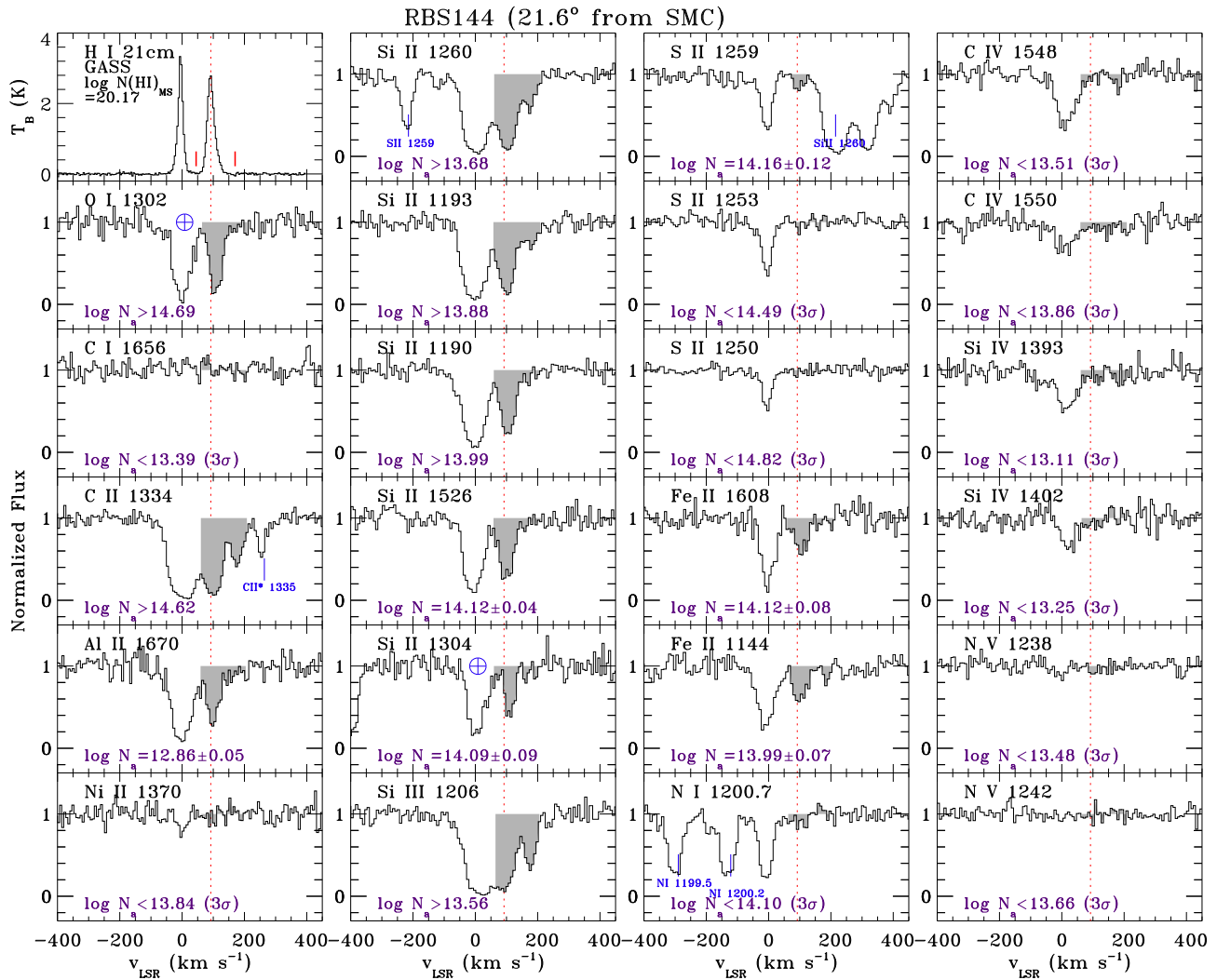} 
\caption{\hst/COS metal-line profiles in the UV spectrum of RBS\,144, 
plus the \hi\ 21\,cm profile from the GASS survey. 
Normalized flux is plotted against LSR velocity for each absorption line. 
Gray shading indicates the MS absorption velocity interval (65--210\kms), 
and the vertical dotted red line shows the central velocity of the MS 21\,cm 
emission. MW absorption in each panel is visible near 0\kms.
Blends are indicated at the appropriate velocity with a tick mark and 
accompanying label. The apparent column density of MS absorption 
is indicated in the lower corner of each panel, with upper limits given
for non-detections. For \oi\ 1302 and \siw\ 1304, 
night-only data are shown to minimize the geocoronal emission.}
\end{figure*}

\begin{figure*}
\epsscale{1.15}
\plotone{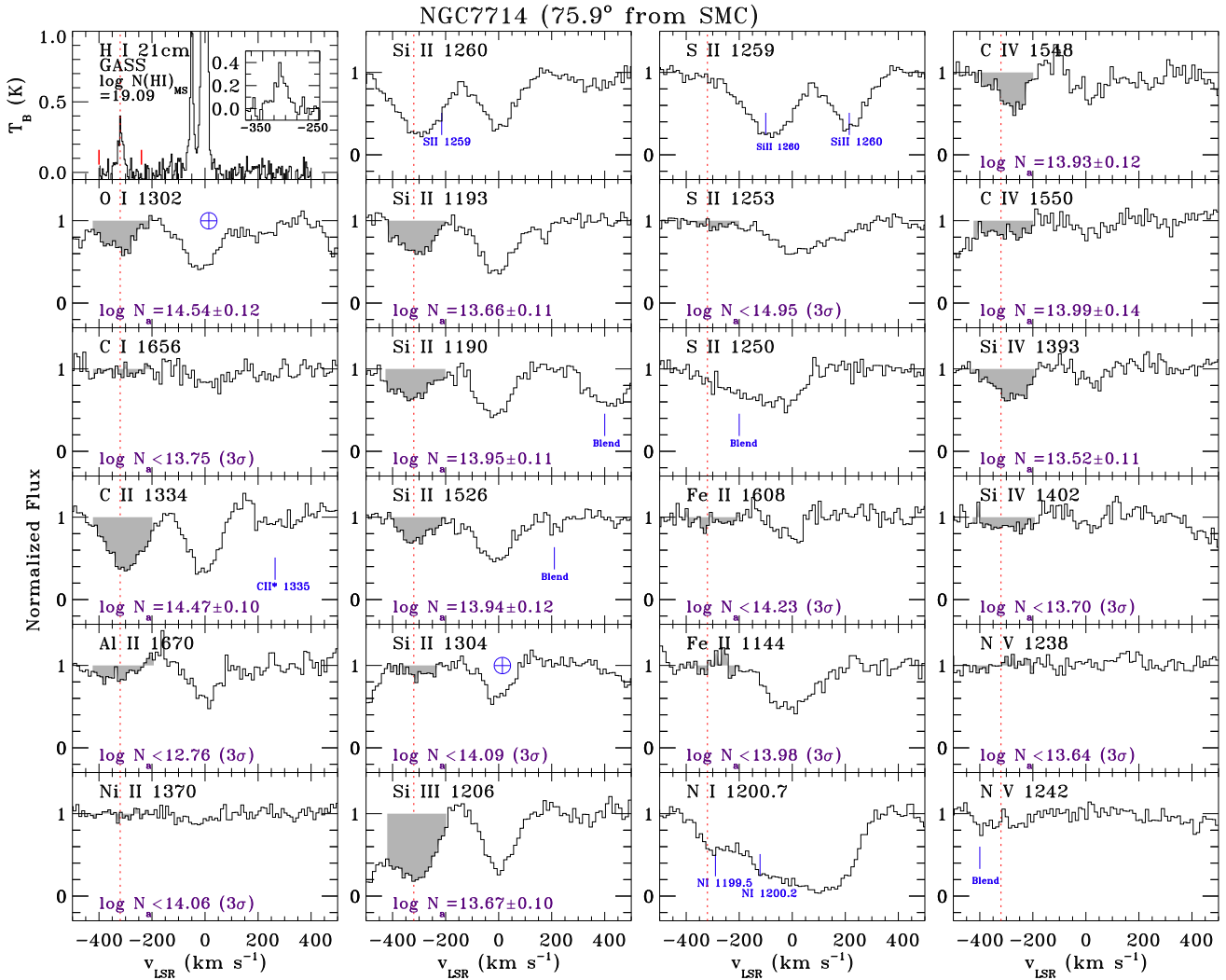} 
\caption{Same as Figure 3, for the NGC\,7714 direction. 
Gray shading indicates MS absorption (--420 to --190\kms),
and the vertical dotted red line shows the central velocity of the 
MS 21\,cm emission. The inset in the 21\,cm panel
shows a zoom-in around the MS component.}
\end{figure*}

\begin{figure*}
\epsscale{1.15}
\plotone{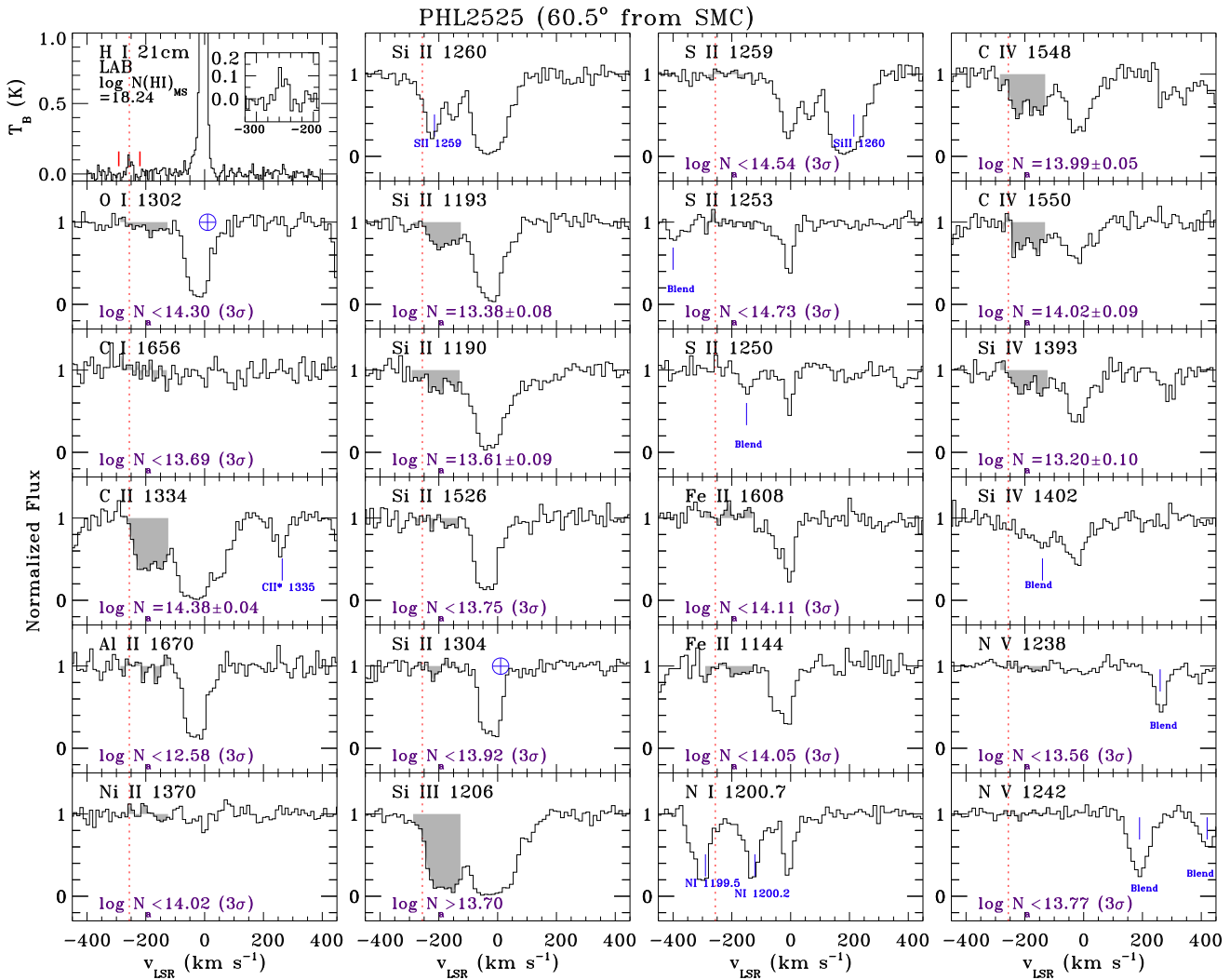} 
\caption{Same as Figure 3, for the PHL\,2525 direction. 
Gray shading indicates MS absorption (--280 to --120\kms),
and the vertical dotted red line shows the central velocity of 
the (weak) MS 21\,cm emission. No \oi\ 1302 absorption
is detected in the MS so only an upper limit on [O/H]$_{\rm MS}$ can
be derived.}
\end{figure*}

\begin{figure*}
\epsscale{1.15}
\plotone{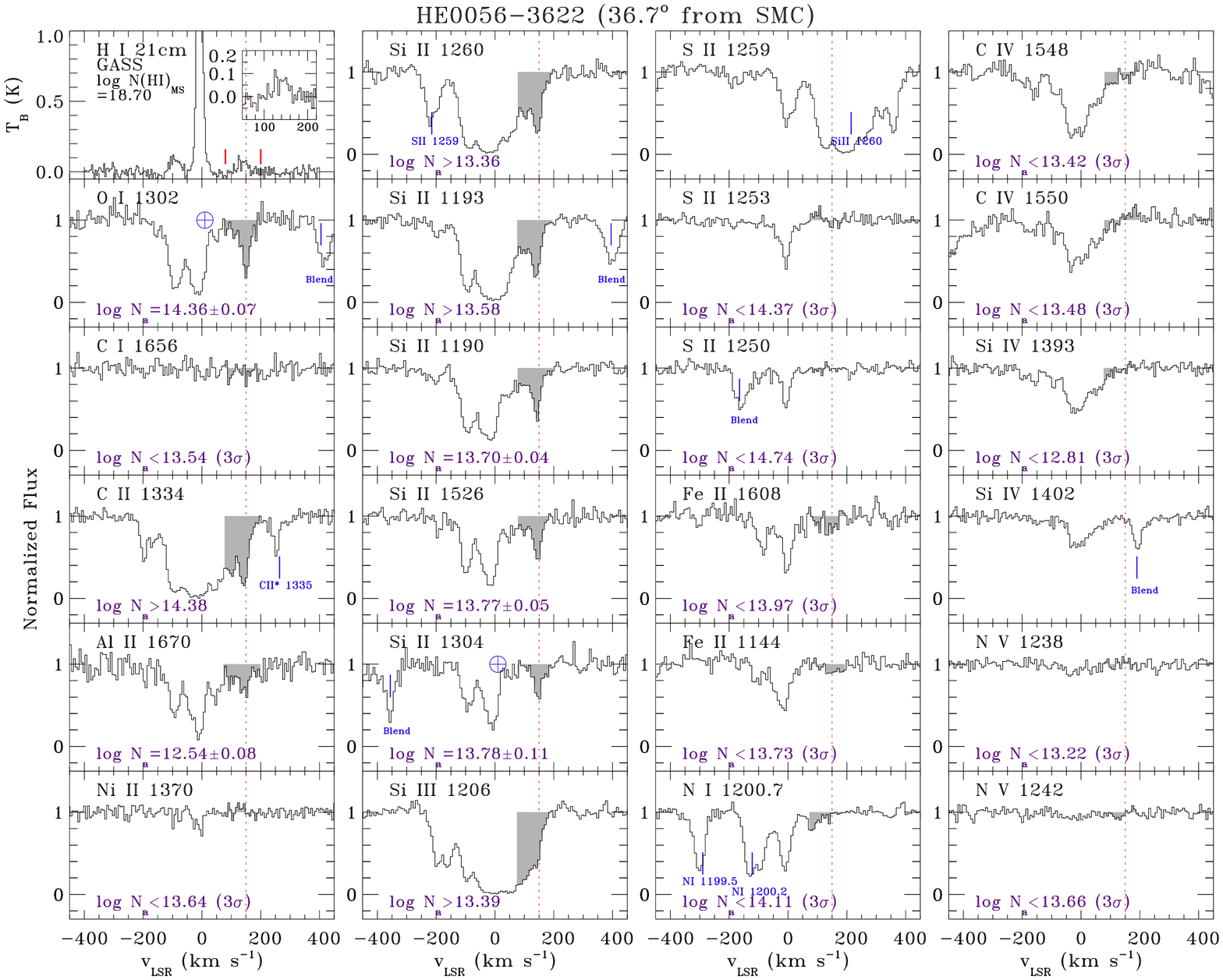} 
\caption{Same as Figure 3, for the HE\,0056--3622 direction. 
Gray shading indicates anomalous-velocity cloud (AVC) absorption 
(80--200\kms). MS absorption in this direction is centered
near --10\kms\ and overlaps with Galactic foreground absorption.
The inset in the 21\,cm panel shows a zoom-in around the AVC component.}
\end{figure*}

\subsection{Measurements}
We used the apparent optical depth (AOD) technique of \cite{SS91} to measure
the absorption in the LSR velocity range of the MS for each metal line of 
interest in the COS and UVES data sets. In this technique,
the AOD in each pixel as calculated as $\tau_a(v)$=ln\,$[F_c(v)/F(v)]$,
where $F_c$ is the continuum level and $F$ is the flux. The 
apparent column density in each pixel is then given by
$N_a(v)= (m_e c/\pi e^2)[\tau_a(v)/f\lambda] = 3.768
\times 10^{14}\,[\tau_a(v)/f\lambda]$ \sqcm\,(km\,s$^{-1}$)$^{-1}$,
where $f$ is the oscillator strength of the transition and $\lambda$ is
the wavelength in Angstroms. This can be integrated over the profile
to give the apparent column density, 
$N_a=\int_{v_{\rm min}}^{v_{\rm max}}N_a(v){\rm d}v$.
The apparent column densities in the MS derived from this technique are 
given in Tables 2, 3, and 4 for the \rbs, \ngc, and \phl\ directions. 
The apparent column densities in the AVC toward \he\ are given in Table 5.

\begin{deluxetable}{lcccc c}
\tabletypesize{\footnotesize}
\tablewidth{0pt}\tabcolsep=2.0pt
\tablecaption{MS Column Densities and Ion Abundances toward RBS144}
\tablehead{Ion & Line & log\,(X/H)$_\odot$\tm{a} & IP\tm{b} & log\,$N_a$(MS)
 & [X$^i$/H]$_{\rm MS}$\tm{c} \\
& (\AA) & & (eV) & ($N_a$ in cm$^{-2}$) &}
\startdata
\hi\ & 21\,cm & 0.0 & 13.6 & 20.17\tm{d} & \nodata \\
  \oi & 1302 & $-$3.31 & 13.6 &              $>$14.69 &                 $>$$-$2.16 \\
  \co & 1656 & $-$3.57 & 11.3 &              $<$13.39 &                 $<$$-$3.20 \\
  \cw & 1334 & $-$3.57 & 24.4 &              $>$14.62 &                 $>$$-$1.98 \\
 \alw & 1670 & $-$5.55 & 18.8 &  12.86$\pm$0.05\tm{e} &     $-$1.75$\pm$0.05\tm{e} \\
 \niw & 1370 & $-$5.78 & 18.2 &              $<$13.84 &                 $<$$-$0.54 \\
 \siw & 1260 & $-$4.49 & 16.3 &              $>$13.68 &                 $>$$-$1.99 \\
 \siw & 1193 & $-$4.49 & 16.3 &              $>$13.88 &                 $>$$-$1.80 \\
 \siw & 1190 & $-$4.49 & 16.3 &              $>$13.99 &                 $>$$-$1.68 \\
 \siw & 1526 & $-$4.49 & 16.3 &        14.12$\pm$0.04 &           $-$1.56$\pm$0.04 \\
 \siw & 1304 & $-$4.49 & 16.3 &        14.09$\pm$0.09 &           $-$1.58$\pm$0.09 \\
 \sit & 1206 & $-$4.49 & 33.5 &              $>$13.56 &                 $>$$-$2.12 \\
  \sw & 1259 & $-$4.88 & 23.3 &  14.16$\pm$0.12\tm{f} &     $-$1.13$\pm$0.12\tm{f} \\
  \sw & 1253 & $-$4.88 & 23.3 &              $<$14.49 &                 $<$$-$0.79 \\
  \sw & 1250 & $-$4.88 & 23.3 &              $<$14.82 &                 $<$$-$0.46 \\
 \few & 1608 & $-$4.50 & 16.2 &        14.12$\pm$0.08 &           $-$1.55$\pm$0.08 \\
 \few & 1144 & $-$4.50 & 16.2 &        13.99$\pm$0.07 &           $-$1.68$\pm$0.07 \\
  \no & 1200.0 & $-$4.17 & 14.5 &              $<$14.10 &                 $<$$-$1.90 \\
  \cf & 1548 & $-$3.57 & 64.5 &              $<$13.51 &                 $<$$-$3.08 \\
  \cf & 1550 & $-$3.57 & 64.5 &              $<$13.86 &                 $<$$-$2.74 \\
 \sif & 1393 & $-$4.49 & 45.1 &              $<$13.11 &                 $<$$-$2.57 \\
 \sif & 1402 & $-$4.49 & 45.1 &              $<$13.25 &                 $<$$-$2.43 \\
  \nf & 1238 & $-$4.17 & 97.9 &              $<$13.48 &                 $<$$-$2.51 \\
  \nf & 1242 & $-$4.17 & 97.9 &              $<$13.66 &                 $<$$-$2.33 \\
\caw & 3934 & $-$5.66 & 11.9 & 12.27$\pm$0.02 & $-$2.23$\pm$0.10\\
\caw & 3969 & $-$5.66 & 11.9 & 12.34$\pm$0.03 & $-$2.16$\pm$0.10\\
\nao & 5891 & $-$5.76 &  5.1 & $<$11.12 &  $<-3.28$\\
\nao & 5897 & $-$5.76 &  5.1 & $<$11.42 &  $<-2.99$\\
\tiw & 3384 & $-$7.05 & 13.6 & $<$11.81 &  $<-1.30$\\
\enddata
\vspace{-1.8cm}
\tablecomments{~LSR velocity integration range for MS in this direction is   65 to  210\kms. For \oi\ 1302 and \siw\ 1304, measurements were made on night-only data.}
\tn{a}{Solar photospheric abundance (Asplund et al. 2009).}\\
\tn{b}{Ionization potential $X^i\to X^{i+1}$.}\\
\tn{c}{[X$^i$/H]$_{\rm MS}$=log\,[X$^i$/\hi]$_{\rm MS}$--log\,(X/H)$_\odot$. Upper/lower limits are 3$\sigma$/1$\sigma$. These abundances are not corrected for ionization. Only the statistical error is given.}\\
\tn{d}{Value given is GASS survey measurement (14.4\arcmin\ beam). LAB survey gives log\,$N$(\hi)=20.27 (30\arcmin\ beam).}\\
\tn{e}{Line potentially saturated.}\\
\tn{f}{Velocity range adjusted to 65--135\kms\ to avoid blend.}\\
\end{deluxetable}
\begin{deluxetable}{lcccc c}
\tabletypesize{\footnotesize}
\tablewidth{0pt}\tabcolsep=2.0pt
\tablecaption{MS Column Densities and Ion Abundances toward NGC7714}
\tablehead{Ion & Line & log\,(X/H)$_\odot$\tm{a} & IP\tm{b} & log\,$N_a$(MS)
 & [X$^i$/H]$_{\rm MS}$\tm{c} \\
& (\AA) & & (eV) & ($N_a$ in cm$^{-2}$) &}
\startdata
\hi\ & 21\,cm & 0.0 & 13.6 & 19.09\tm{d} & \nodata \\
  \oi & 1302 & $-$3.31 & 13.6 &        14.54$\pm$0.12 &           $-$1.24$\pm$0.12 \\
  \co & 1656 & $-$3.57 & 11.3 &              $<$13.75 &                 $<$$-$1.77 \\
  \cw & 1334 & $-$3.57 & 24.4 &        14.47$\pm$0.10 &           $-$1.05$\pm$0.10 \\
 \alw & 1670 & $-$5.55 & 18.8 &              $<$12.76 &                 $<$$-$0.78 \\
 \niw & 1370 & $-$5.78 & 18.2 &              $<$14.06 &                    $<$0.75 \\
 \siw & 1193 & $-$4.49 & 16.3 &        13.66$\pm$0.11 &           $-$0.94$\pm$0.11 \\
 \siw & 1190 & $-$4.49 & 16.3 &        13.95$\pm$0.11 &           $-$0.65$\pm$0.11 \\
 \siw & 1526 & $-$4.49 & 16.3 &        13.94$\pm$0.12 &           $-$0.66$\pm$0.12 \\
 \siw & 1304 & $-$4.49 & 16.3 &              $<$14.09 &                 $<$$-$0.51 \\
 \sit & 1206 & $-$4.49 & 33.5 &        13.67$\pm$0.10 &           $-$0.93$\pm$0.10 \\
  \sw & 1253 & $-$4.88 & 23.3 &              $<$14.95 &                    $<$0.74 \\
 \few & 1608 & $-$4.50 & 16.2 &              $<$14.23 &                 $<$$-$0.36 \\
 \few & 1144 & $-$4.50 & 16.2 &              $<$13.98 &                 $<$$-$0.61 \\
  \cf & 1548 & $-$3.57 & 64.5 &        13.93$\pm$0.12 &           $-$1.59$\pm$0.12 \\
  \cf & 1550 & $-$3.57 & 64.5 &        13.99$\pm$0.14 &           $-$1.53$\pm$0.14 \\
 \sif & 1393 & $-$4.49 & 45.1 &        13.52$\pm$0.11 &           $-$1.08$\pm$0.11 \\
 \sif & 1402 & $-$4.49 & 45.1 &              $<$13.70 &                 $<$$-$0.90 \\
  \nf & 1238 & $-$4.17 & 97.9 &              $<$13.64 &                 $<$$-$1.28 \\
\enddata
\vspace{-1.2cm}
\tablecomments{~LSR velocity integration range for MS in this direction is -420 to -190\kms. For \oi\ 1302 and \siw\ 1304, measurements were made on night-only data.}
\tn{a}{Solar photospheric abundance (Asplund et al. 2009).}\\
\tn{b}{Ionization potential $X^i\to X^{i+1}$.}\\
\tn{c}{[X$^i$/H]$_{\rm MS}$=log\,[X$^i$/\hi]$_{\rm MS}$--log\,(X/H)$_\odot$. Upper/lower limits are 3$\sigma$/1$\sigma$. These abundances are not corrected for ionization. Only the statistical error is given.}\\
\tn{d}{Value given is GASS survey measurement (14.4\arcmin\ beam).}\\
\end{deluxetable}
\begin{deluxetable}{lcccc c}
\tabletypesize{\footnotesize}
\tablewidth{0pt}\tabcolsep=2.0pt
\tablecaption{MS Column Densities and Ion Abundances toward PHL2525}
\tablehead{Ion & Line & log\,(X/H)$_\odot$\tm{a} & IP\tm{b} & log\,$N_a$(MS)
 & [X$^i$/H]$_{\rm MS}$\tm{c} \\
& (\AA) & & (eV) & ($N_a$ in cm$^{-2}$) &}
\startdata
\hi\ & 21\,cm & 0.0 & 13.6 & 18.24\tm{d} & \nodata \\
  \oi & 1302 & $-$3.31 & 13.6 &              $<$14.30 &                 $<$$-$0.63 \\
  \co & 1656 & $-$3.57 & 11.3 &              $<$13.69 &                 $<$$-$0.98 \\
  \cw & 1334 & $-$3.57 & 24.4 &        14.38$\pm$0.04 &           $-$0.29$\pm$0.04 \\
 \alw & 1670 & $-$5.55 & 18.8 &              $<$12.58 &                 $<$$-$0.11 \\
 \niw & 1370 & $-$5.78 & 18.2 &              $<$14.02 &                    $<$1.56 \\
 \siw & 1193 & $-$4.49 & 16.3 &        13.38$\pm$0.08 &           $-$0.37$\pm$0.08 \\
 \siw & 1190 & $-$4.49 & 16.3 &        13.61$\pm$0.09 &           $-$0.14$\pm$0.09 \\
 \siw & 1526 & $-$4.49 & 16.3 &              $<$13.75 &                 $<$$-$0.00 \\
 \siw & 1304 & $-$4.49 & 16.3 &              $<$13.92 &                    $<$0.17 \\
 \sit & 1206 & $-$4.49 & 33.5 &              $>$13.70 &                 $>$$-$0.05 \\
  \sw & 1259 & $-$4.88 & 23.3 &              $<$14.54 &                    $<$1.18 \\
  \sw & 1253 & $-$4.88 & 23.3 &              $<$14.73 &                    $<$1.37 \\
 \few & 1608 & $-$4.50 & 16.2 &              $<$14.11 &                    $<$0.37 \\
 \few & 1144 & $-$4.50 & 16.2 &              $<$14.05 &                    $<$0.30 \\
  \cf & 1548 & $-$3.57 & 64.5 &        13.99$\pm$0.05 &           $-$0.68$\pm$0.05 \\
  \cf & 1550 & $-$3.57 & 64.5 &        14.02$\pm$0.09 &           $-$0.65$\pm$0.09 \\
 \sif & 1393 & $-$4.49 & 45.1 &        13.20$\pm$0.10 &           $-$0.55$\pm$0.10 \\
  \nf & 1238 & $-$4.17 & 97.9 &              $<$13.56 &                 $<$$-$0.52 \\
  \nf & 1242 & $-$4.17 & 97.9 &              $<$13.77 &                 $<$$-$0.30 \\
\enddata
\vspace{-1.2cm}
\tablecomments{~LSR velocity integration range for MS in this direction is -280 to -120\kms. For \oi\ 1302 and \siw\ 1304, measurements were made on night-only data.}
\tn{a}{Solar photospheric abundance (Asplund et al. 2009).}\\
\tn{b}{Ionization potential $X^i\to X^{i+1}$.}\\
\tn{c}{[X$^i$/H]$_{\rm MS}$=log\,[X$^i$/\hi]$_{\rm MS}$--log\,(X/H)$_\odot$. Upper/lower limits are 3$\sigma$/1$\sigma$. These abundances are not corrected for ionization. Only the statistical error is given.}\\
\tn{d}{Value given is GASS survey measurement (14.4\arcmin\ beam).}\\
\end{deluxetable}
\begin{deluxetable}{lcccc c}
\tabletypesize{\footnotesize}
\tablewidth{0pt}\tabcolsep=2.0pt
\tablecaption{AVC Column Densities and Ion Abundances toward HE0056-3622}
\tablehead{Ion & Line & log\,(X/H)$_\odot$\tm{a} & IP\tm{b} & log\,$N_a$(MS)
 & [X$^i$/H]$_{\rm MS}$\tm{c} \\
& (\AA) & & (eV) & ($N_a$ in cm$^{-2}$) &}
\startdata
\hi\ & 21\,cm & 0.0 & 13.6 & 18.70\tm{d} & \nodata \\
  \oi & 1302 & $-$3.31 & 13.6 &        14.36$\pm$0.07 &           $-$1.03$\pm$0.07 \\
  \co & 1656 & $-$3.57 & 11.3 &              $<$13.54 &                 $<$$-$1.59 \\
  \cw & 1334 & $-$3.57 & 24.4 &              $>$14.38 &                 $>$$-$0.75 \\
 \alw & 1670 & $-$5.55 & 18.8 &        12.54$\pm$0.08 &           $-$0.61$\pm$0.08 \\
 \niw & 1370 & $-$5.78 & 18.2 &              $<$13.64 &                    $<$0.72 \\
 \siw & 1260 & $-$4.49 & 16.3 &              $>$13.36 &                 $>$$-$0.85 \\
 \siw & 1193 & $-$4.49 & 16.3 &              $>$13.58 &                 $>$$-$0.63 \\
 \siw & 1190 & $-$4.49 & 16.3 &        13.70$\pm$0.04 &           $-$0.51$\pm$0.04 \\
 \siw & 1526 & $-$4.49 & 16.3 &        13.77$\pm$0.05 &           $-$0.44$\pm$0.05 \\
 \siw & 1304 & $-$4.49 & 16.3 &        13.78$\pm$0.11 &           $-$0.42$\pm$0.11 \\
 \sit & 1206 & $-$4.49 & 33.5 &              $>$13.39 &                 $>$$-$0.82 \\
  \sw & 1253 & $-$4.88 & 23.3 &              $<$14.37 &                    $<$0.55 \\
  \sw & 1250 & $-$4.88 & 23.3 &              $<$14.74 &                    $<$0.92 \\
 \few & 1608 & $-$4.50 & 16.2 &              $<$13.97 &                 $<$$-$0.23 \\
 \few & 1144 & $-$4.50 & 16.2 &              $<$13.73 &                 $<$$-$0.47 \\
  \no & 1200.0 & $-$4.17 & 14.5 &              $<$14.11 &                 $<$$-$0.42 \\
  \cf & 1548 & $-$3.57 & 64.5 &              $<$13.42 &                 $<$$-$1.71 \\
  \cf & 1550 & $-$3.57 & 64.5 &              $<$13.48 &                 $<$$-$1.65 \\
 \sif & 1393 & $-$4.49 & 45.1 &              $<$12.81 &                 $<$$-$1.40 \\
  \nf & 1238 & $-$4.17 & 97.9 &              $<$13.22 &                 $<$$-$1.31 \\
  \nf & 1242 & $-$4.17 & 97.9 &              $<$13.66 &                 $<$$-$0.86 \\
\caw & 3934 & $-$5.66 & 11.9 & 11.99$\pm$0.04 & $-$1.05$\pm$0.11\\
\caw & 3969 & $-$5.66 & 11.9 & 12.13$\pm$0.06 & $-$0.91$\pm$0.12\\
\nao & 5891 & $-$5.76 &  5.1 & $<$11.37 & $<-1.57$\\
\nao & 5897 & $-$5.76 &  5.1 & $<$11.63 & $<-1.31$\\
\tiw & 3384 & $-$7.05 & 13.6 & $<$11.94 & $<$0.30\\
\enddata
\vspace{-1.2cm}
\tablecomments{~LSR velocity integration range for AVC in this direction is   80 to  200\kms. For \oi\ 1302 and \siw\ 1304, measurements were made on night-only data.}
\tn{a}{Solar photospheric abundance (Asplund et al. 2009).}\\
\tn{b}{Ionization potential $X^i\to X^{i+1}$.}\\
\tn{c}{[X$^i$/H]$_{\rm MS}$=log\,[X$^i$/\hi]$_{\rm MS}$--log\,(X/H)$_\odot$. Upper/lower limits are 3$\sigma$/1$\sigma$. These abundances are not corrected for ionization. Only the statistical error is given.}\\
\tn{d}{Value given is GASS survey measurement (14.4\arcmin\ beam). LAB survey gives log\,$N$(\hi)=18.87 (30\arcmin\ beam).}\\
\end{deluxetable}

All our data have sufficient S/N (see Table 1) to lie in the regime where 
the AOD method is reliable \citep{Fo05b}, 
so long as no unresolved saturation is present.
This could occur if narrow unresolved lines are present in the profiles,
which is more likely for the COS data than the UVES data.
In cases where doublets or multiplets of the same ion are present 
in the data (e.g. \siw), we used these to check for saturation.
When lines are not detected at 3$\sigma$ significance, we give upper limits
on $N_a$ based on the noise measured in the continuum.

For the UVES data, Voigt-component fits to the optical (\caw, \tiw, and 
\nao, where present) lines were used to determine the low-ion component 
structure in the Stream, using the {\sc VPFIT} software 
package\footnotemark[3]\footnotetext[3]{Available at
http://www.ast.cam.ac.uk/$\sim$rfc/vpfit.html.}.
These fits account for the UVES instrumental line spread function, assumed
to be a Gaussian with a FWHM of 4.3\kms.
Each ion was fit independently, with the number of components
chosen to be the minimum necessary to match the data (see Figure 2):
three MS components for \rbs\ and two AVC components for \he.
Note that we use the UVES data to derive the component structure 
because of their superior velocity resolution compared to the COS data 
(4.3\kms\ vs $\approx$19\kms\ FWHM), though we note that \caw\ does not 
necessarily trace the same low-ion phase as the UV metal lines because 
of its low ionization potential.

\subsection{MS Absorption toward RBS\,144}
The central LSR velocity of the MS in the \rbs\ direction is 92\kms,
as defined by the peak of the GASS 21\,cm emission profile.
The \hi\ profile extends over the range 75--140\kms.
The UV absorption lines show absorption centered near the same velocity, 
but covering a more extended interval of 65--210\kms\ (Figure 3).

The following UV metal lines are detected in the MS: 
\oi\ 1302, \cw\ 1334, \siw\ 1260,1193,1190,1526,1304, 
\sit\ 1206, \sw\ 1259, \alw\ 1670, and \few\ 1144,1608.
Several of these lines are saturated; the lines which appear unsaturated, 
from which we derive ionic column densities,
are \siw\ 1526,1304, \sw\ 1259, \alw\ 1670, and \few\ 1144,1608
(see Table 2). However, for \alw\ 1670, we cannot rule out 
unresolved saturation, since the line reaches a normalized depth of 0.8.
No detection is seen in the high ion doublets
\cf\ 1548,1550, \sif\ 1393,1402, and \nf\ 1238,1242,
or in the low-ion lines \nit\ 1370, \sw\ 1253,1250, or \no\ 1200.7.
Among the \no\ 1199.5, 1200.2, 1200.7 triplet, only the 1200.7 line is 
shown on Figure 3 since absorption at MS velocities in the first two 
lines is blended. 
Two high-velocity components can be discerned in the UV profiles.
The primary MS component (closely aligned with the 21\,cm emission) 
is centered near 100\kms.
A secondary component at 180\kms\ is seen in \cw, \siw, and \sit,
but not in \oi\ or \sw, or \hi\ 21\,cm (Figure 3).

The VLT/UVES data give MS detections of \caw\ 3934,3969 and \tiw\ 3384 
(low significance), and upper limits on \nao\ 5891,5897.
Three components of MS absorption are visible in the \caw\ profile, spread
over 25\kms, a much narrower interval than the UV metal lines, with narrow
$b$-values of 6.1$\pm$0.4\kms, 3.0$\pm$1.6\kms, and 6.2$\pm$0.5\kms.
These lines trace the cool-gas component structure in the Stream (see Paper II).
An offset of 6\kms\ is seen between the velocity centroids of the \hi\ 
and \caw\ MS components (see Figure 2):
the \hi\ MS emission is centered at 92\kms\ (in both the LAB and the GASS 
data), whereas the \caw\ absorption is centered at 98$\pm$1\kms. Furthermore, 
the two higher-velocity components seen in \caw\ at 107$\pm$1\kms\ and 
120$\pm$1\kms\ show no analogues in 
the 21\,cm data. This indicates that small-scale structure exists in the 
pencil-beam sightline that is not resolved in the 21\,cm beam.

However, since the \hi\ column in the MS in this direction is high enough 
to contribute to the damping wings on the Galactic \lya\ line, we can
derive $N$(\hi)$_{\rm MS}$ {\it in the pencil-beam line-of-sight}
by fitting the observed \lya\ profile with a two-component (MW+MS) model, 
following \citet{Le08}.
This is important since it eliminates the systematic error on metallicities
that derives from comparing UV lines measured with an infinitesimal 
beam with 21\,cm \hi\ measurements made with a finite beam.
We fix the velocity components at
--5$\pm$5\kms\ (MW component) and 95$\pm$5\kms\ (MS component)
to follow the component structure seen in the UV metal lines,
and allow only the \hi\ column density in each component to vary.
Using a fourth-order polynomial fit to the continuum, 
we find the \lya\ profile is well reproduced ($\chi_\nu^2$=0.88) by a model with
log\,$N$(\hi)$_{\rm MW}$=20.00$\pm$0.05 and
log\,$N$(\hi)$_{\rm MS}$=20.09$\pm$0.04 (see Figure 7). 
The latter column matches the GASS value 
log\,$N$(\hi)$_{\rm MS}$=20.17 within 2$\sigma$, illustrating that in this 
direction the systematic uncertainty on log\,$N$(\hi)$_{\rm MS}$ due to 
beamsize effects is small ($\la$0.1\,dex). 
The accuracy of this technique is limited by the true component structure; 
if additional components are present, the relative columns in the MW and MS 
components would change. Further discussion of these limitations is given 
in \citet{Le08}.
This technique cannot be used to derive $N$(\hi)$_{\rm MS}$
along the other three sightlines in this paper since in those directions 
$N$(\hi)$_{\rm MW}\!\gg\!N$(\hi)$_{\rm MS}$.

\begin{figure}
\epsscale{1.15}
\plotone{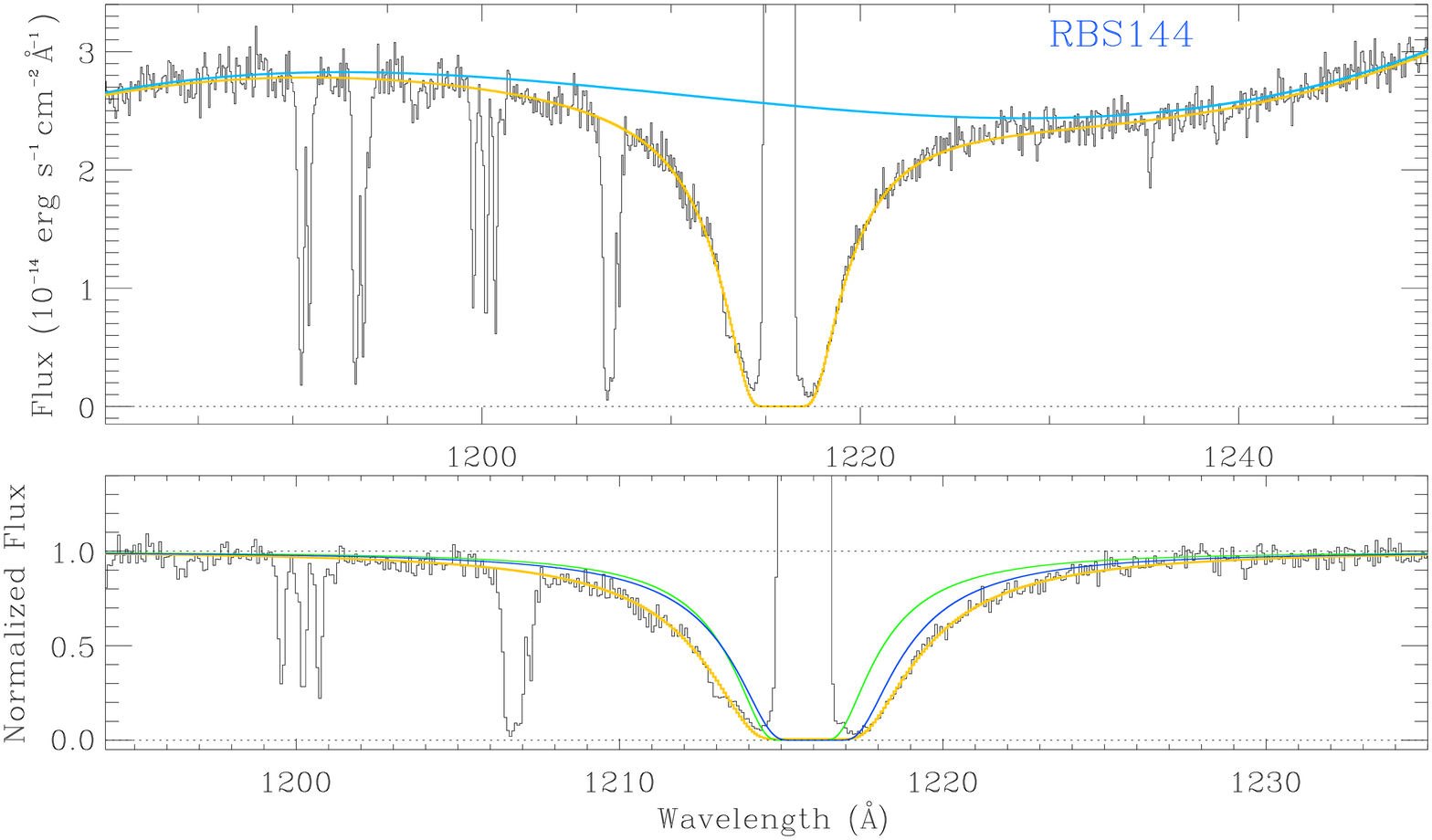}
\caption{Derivation of $N$(\hi) toward RBS\,144 from a two-component 
(MW+MS) fit to the damping wings of \lya\ (shown in yellow), using a 
fourth-order polynomial fit to the continuum (shown in blue).
The upper and lower panels show the raw and normalized profiles, respectively.
The strong emission feature near the center of each panel is geocoronal \lya\
emission. The fit yields log\,$N$(\hi)$_{\rm MW}$=20.00$\pm$0.05 and 
log\,$N$(\hi)$_{\rm MS}$=20.09$\pm$0.04, which are in reasonable agreement 
with the columns derived from the GASS 21\,cm data.
The contributions from the two components are shown in the lower
panel as green (MW) and blue (MS) curves.
These COS data have been binned to the instrumental resolution of 18\kms.}
\end{figure}

\subsection{MS Absorption toward NGC\,7714}
\ngc\ is an On-Stream direction showing a well-defined 21\,cm emission component
at --320\kms\ with log\,$N$(\hi)$_{\rm MS}$=19.09 measured from the GASS data
(18.93 measured from the LAB data), and UV absorption 
centered at the same velocity but extending over a wider interval of
--420 to --190\kms\ (Figure 4). The UV lines detected in the MS are \oi\ 
1302, \cw\ 1334, \siw\ 1193,1190,1526, \sit\ 1206 and the high-ion lines 
\cf\ 1548,1550 and \sif\ 1393. The component structure in the UV lines is 
fairly simple, with a single broad component well aligned with the 21\,cm 
emission. Since this target is slightly extended (a starburst nucleus 
rather than a point source), the absorption is not as well defined as for the
other sightlines in our sample. No UVES data are available for this sightline.
Ionic column densities are given in Table 3.

\subsection{MS Absorption toward PHL\,2525}
The LAB 21\,cm data toward \phl\ (30\arcmin\ beam) show a very weak emission 
component log\,$N$(\hi)$_{\rm MS}$=18.24 at --260\kms\ (see Figure 5, top-left 
panel). We identify this component with the MS.
This component is not visible in the GASS data (14.4\arcmin\ beam), 
which give log\,$N$(\hi)$_{\rm MS}\!<\!18.21$ (3$\sigma$), 
marginally inconsistent with the LAB data.
UV absorption covering the interval --300 to --100\kms\ is seen in \cw\ 1334, 
\siw\ 1193,1190, \sit\ 1206, \cf\ 1548,1500, and \sif\ 1393,1402 (Figure 5). 
No absorption in this interval is seen in the UVES data covering 
\caw, \nao, and \tiw\ (Figure 2).
There is a clear offset of $\approx$50\kms\ between the center of the UV 
absorption and the LAB \hi\ emission, further indicating the beamsize mismatch.
Because of this issue, this sightline is of little use for deriving robust 
abundances. However, the combination of the non-detection of \oi\ 1302 in the 
MS velocity interval with the strength of the LAB \hi\ emission allows us to 
place an upper limit on the MS metallicity (see \S4). Ionic column densities 
for this sightline are given in Table 4.

\subsection{AVC Absorption toward HE\,0056--3622}
In this direction high-velocity absorption is detected in many 
UV lines centered at 150\kms\ and covering the interval 80--200\kms\ (Figure 6).
As discussed in \S1, this absorption appears to trace the 
AVCs found near the South Galactic Pole \citep{Pu03a}.
These AVCs may have an origin with the Sculptor group of galaxies 
\citep{Ma75}, whose velocities are are low as 70\kms, or may represent
fragments of the MS \citep{HR79} that have separated kinematically from 
the principal filaments.
The 21\,cm emission component from the AVC is weak, with 
log\,$N$(\hi)$_{\rm AVC}$=18.70 measured over the interval 80--200\kms\ 
in the GASS data. 
The AVC is detected in absorption in \oi\ 1302, \cw\ 1334, 
\siw\ 1260,1193,1190,1526,1304, \sit\ 1206, \alw\ 1670, and \caw\ 3934,3969, 
but not in \sw\ 1253,1250, \few\ 1144,1608, \cf\ 1548,1550, \sif\ 1393,1402, 
\nao\ 5891,5897, or \tiw\ 3384. 
Among the detected lines, \oi\ 1302, \alw\ 1670, \siw\ 1190,1526,1304,
and \caw\ 3934,3969 are unsaturated and can be used to determine the 
ionic column densities (Table 5).

We fit two closely-spaced components to the \caw\ absorption centered
near 150\kms\ (Figure 2), but the \caw\ absorption does not cover
the full velocity interval of the AVC absorption seen in the UV lines,
which show one component centered at 100\kms, and a second weaker 
component seen in \sit\ 1206 and \siw\ 1260 at 170\kms.

\section{Chemical Abundances in the MS}
The UV data contain a rich variety of diagnostics on elemental abundances 
in the Stream. Among the available abundance indicators, 
the \oi/\hi\ and \sw/\hi\ ratios are considered the most reliable,
being the least affected by dust and ionization effects 
\citep[][]{FS71, SS96, Me98, Je05, Je09}.
However, at the high \hi\ columns [log\,$N$(\hi)$\approx$20] and 
low metallicities (O/H$\approx$0.1 solar) found in the MS, 
\oi\ 1302 saturates yet \oi\ 1356 is too weak to detect,
so only a lower limit on the \oi\ column is measurable.
In this regime, \sw/\hi\ is the better metallicity indicator.
Thus the metallicity indicator we use in each sightline depends on
the \hi\ column.

Toward \rbs, we measure log\,$N_{\rm a}$(\oi)$>$14.69 from 
the strongly saturated \oi\ 1302 line, giving a non-constraining limit 
[\oi/\hi]$_{\rm MS}\!>\!-2.17$. Turning to \sw/\hi, we 
use the \sw\ column measured from an AOD integration of \sw\ 1259
to derive [\sw/\hi]$_{\rm MS}$=$-$1.13$\pm$0.12(stat)$\pm$0.10(syst), 
where the first error is the statistical uncertainty and the second error 
represents the difference between the LAB and GASS \hi\ column densities
and accounts for the beamsize mismatch between the radio and UV observations
(see F10 for a more detailed discussion of the size of this systematic error).
Adding the two errors in quadrature to give an overall uncertainty gives
[\sw/\hi]$_{\rm MS}$=$-$1.13$\pm$0.16. 
The \sw\ 1259 line is only detected at 3.3$\sigma$ significance in the co-added 
spectrum (Figure 3), but is seen in both individual sub-exposures.
A non-detection would serve to lower the derived metallicity.
For this \sw\ 1259 line only, we adjusted the velocity integration range to
65--135\kms\ instead of 65--210\kms, to avoid blending with Galactic \siw\ 1260.
Fortunately the 21\,cm MS signal is almost all contained in this restricted 
velocity range, so the S abundance is not affected by this 
adjustment. Using the non-detection of MS absorption in \no\ 1200.7, we derive 
log\,$N$(\no)$<$14.10 and [\no/\hi]$_{\rm MS}\!<\!-1.90$ (3$\sigma$). 
Nitrogen is therefore underabundant in the gas phase of the Stream, with 
[N/S]$_{\rm MS}$=[\no/\sw]$_{\rm MS}\!<\!-0.77$ (3$\sigma$). 

Toward \ngc, we measure log\,$N_{\rm a}$(\oi)=14.54$\pm$0.06 in the MS from 
the night-only \oi\ 1302 data, corresponding
to [\oi/\hi]$_{\rm MS}$=$-$1.24$\pm$0.12(stat)$\pm$0.16(syst)=$-$1.24$\pm$0.20 
when combining the errors in quadrature.
Toward \phl, we measure log\,$N_{\rm a}$(\oi)$<$14.30 (3$\sigma$) in the MS, 
again from night-only \oi\ 1302 data, corresponding to 
[\oi/\hi]$_{\rm MS}\!<\!-$0.63 (3$\sigma$). 
Toward \he, we measure log\,$N_{\rm a}$(\oi)=14.36$\pm$0.07 in the AVC from 
the night-only \oi\ 1302 data, corresponding
to [\oi/\hi]$_{\rm AVC}$=$-$1.03$\pm$0.07(stat)$\pm$0.17(syst)=$-$1.03$\pm$0.18.
The \oi\ 1302 line in the AVC appears unsaturated (Figure 6), with a maximum 
normalized depth of 0.7, although we cannot rule out the possibility of 
unresolved saturation; 
if present, this would raise the value of O/H in the AVC.
An upper limit on the \sw\ abundance measured from the non-detection of 
\sw\ 1253 toward \he\ is [\sw/\hi]$_{\rm AVC}<$+0.55 (3$\sigma$), consistent 
with the \oi/\hi\ abundance, but not constraining. 

\subsection{Ion Abundances}
In this sub-section we focus on the low-ion (singly ionized) species and their 
relative proportion with \hi. These ratios form empirical indicators of the 
gas-phase abundances. They are listed in Tables 2, 3, 4, and 5 and are plotted 
in Figure 8 for ten low-ionization species, in order of increasing atomic 
number.

The top panel of Figure 8 compares the MS ion abundances 
measured toward \rbs\ (this paper; log\,$N$(\hi)$_{\rm MS}$=20.17) 
and \fn\ (Paper II; log\,$N$(\hi)$_{\rm MS}$=19.95)
with the compilation of LMC and SMC reference abundances 
presented by \citet{RD92}, which
includes abundances measured in stars and in interstellar 
emission-line objects such as supernova remnants and \hw\ regions.
Comparing the \rbs\ and \fn\ data points, a clear pattern is seen, 
in which the ion abundances 
(or limits) are lower toward \rbs\ than toward \fn\ for all ions shown.
This indicates that as the \hi\ column goes down, the 
low-ion metal columns do not decrease in linear proportion.
The MS ion abundances measured toward RBS\,144 are
consistently lower than the current-day elemental abundances in the LMC and 
SMC. The difference is smallest for sulfur, which shows a 0.6\,dex 
difference between the MS and the current-day SMC value. 
The Magellanic Cloud abundances vary substantially from one element to another, 
which complicates the interpretation of the MS abundance pattern.

In the lower panel of Figure 8, we compare the dust depletions of each 
element measured toward \rbs\ and \fn\ 
with those measured in the Magellanic Bridge by 
\citet{Le01} and \citet{Le02}, and to the compilation of
Milky Way halo depletions presented by \citet{We97}, 
which was based on \citet{SS96} and \citet{Fi96}. 
The Bridge measurements are made in the direction of O-star 
\object{DI\,1388}, whereas the halo cloud measurements represent the 
averages over a sample of many sightlines.
We define the depletion of element X relative to sulfur as
$\delta$(X)$\equiv$[X/\sw]=[X/H]--[\sw/H]. 
The depletions measured in the two MS sightlines are generally similar, 
with a notable discrepancy for \caw: $\delta$(Ca) is 0.44\,dex lower toward 
\fn\ than toward \rbs. 
The Bridge and Stream show a depletion pattern that is consistent for all 
ions shown, particularly for $\delta$(Fe) and $\delta$(Si), where values 
rather than limits are available. The MS dust depletion pattern toward 
\rbs\ is also consistent with the Galactic halo depletion pattern 
for all elements shown except Nitrogen, which is underabundant in the 
MS in this direction relative to the Galactic halo by $\ga$0.7\,dex
(although the halo value is highly uncertain).

\begin{figure*}  
\epsscale{1.15}
\plotone{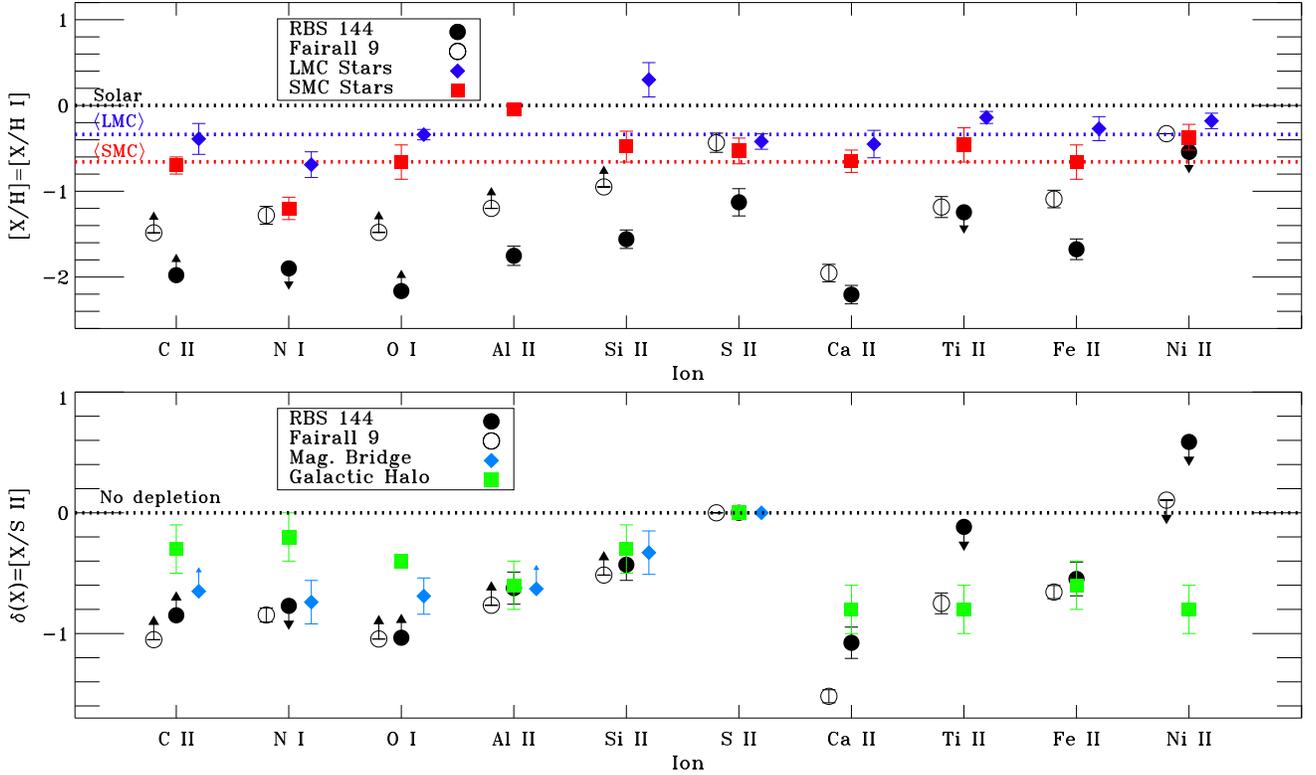}   
\caption{{\bf Upper panel}: comparison of MS ion abundances measured toward
RBS\,144 (this paper) and Fairall\,9 (Paper II) with the SMC and LMC abundance 
patterns presented by \citet{RD92}. 
The solar, average LMC (0.46 solar), and average SMC (0.22 solar) 
abundances are plotted as horizontal lines.
Consistently lower ion abundances are seen toward 
RBS\,144 [log\,$N$(\hi)$_{\rm MS}$=20.17] than toward Fairall\,9
[log\,$N$(\hi)$_{\rm MS}$=19.95].
{\bf Lower panel}: comparison of the dust depletion levels 
$\delta$(X)=[X/\sw] measured in the MS toward RBS\,144 and Fairall\,9 
with the dust depletion pattern measured 
in the Magellanic Bridge by \citet{Le02}, and with the Galactic halo 
depletion compilation of \citet{We97}. All values are uncorrected for 
ionization. The data points have been offset 
in the x-direction for clarity. See \S4.1 for discussion.}
\end{figure*}

\subsection{Gas-to-dust ratios} 
A useful number to derive when discussing the chemical properties of the
Stream is the gas-to-dust mass ratio, which is directly
related to the depletion of Fe atoms onto dust grains.
We define this ratio following \citet{Wo03} as
G/D\,$\equiv N$(H)/$N$(Fe)$_{\rm d}$
where $N$(Fe)$_d$=$N$(S)(Fe/S)$_\odot$(1--10$^{\rm [Fe/S]}$)
is the column of iron in dust grains. This assumes that
a solar Fe/S ratio applies to the entire (dust+gas) system.
The single sightline in our sample with both \few\ and \sw\ directions
(and hence for which we can calculate G/D)
is \rbs; in this direction, we derive 
(G/D)$_{\rm MS}$=$600000_{-190000}^{+320000}$. 
We can normalize this to the Galactic ISM value by writing
(G/D)$_{\rm norm}$=(G/D)$_{\rm MS}$/(G/D)$_{\rm MW}$. Since  
(G/D)$_{\rm MW}\!\approx$(Fe/H)$_\odot^{-1}$ as Fe is almost entirely depleted 
in the Galactic ISM, we find (G/D)$_{\rm norm}$=19$^{+10}_{-6}$.
Along the nearby \fn\ sightline discussed in Paper II,
we find a lower ratio (G/D)$_{\rm MS}$=$104000_{-14000}^{+17000}$ 
and (G/D)$_{\rm norm}$=3.3$^{+0.5}_{-0.5}$.
This corresponds to a higher dust content toward \fn, as expected given the 
higher MS metallicity in that direction.

For comparison, measurements of (G/D)$_{\rm norm}$ in the SMC
lie in the range $\approx$5--14 
\citep[][J. Roman-Duval et al. 2013, in prep.]{So06, Le07, Go09}, 
whereas the LMC has a lower average ratio
(G/D)$_{\rm norm}\!\approx$2--4 \citep{Dr03}.
The ratio observed in the MS toward \rbs\ is therefore
closer to (though higher than) the current-day ratio observed in the diffuse 
gas of the SMC, whereas the ratio measured toward \fn\ is closer 
to the LMC value. 

\subsection{Ionization Corrections}
Up to now we have considered \emph{ion} abundances, not
\emph{elemental} abundances. 
The abundance of the species \oi\ and \sw\ are robust indicators of the 
elemental abundances of O and Si, as discussed above, but for other species, 
ionization corrections must be applied to derive the elemental abundances.
These corrections can be derived under the standard assumption that the 
low-ionization species (singly and doubly ionized)
are photoionized by the incident radiation field. 
We ran a set of photoionization models using 
\emph{Cloudy} v10.00 \citep{Fe98} to investigate this process, 
assuming the gas exists in a plane-parallel slab of uniform density.
We follow the two-step method outlined in F10. 
First, we solve for the value of the ionization parameter 
log\,$U\!\equiv$\,log\,($n_\gamma/n_{\rm H}$) necessary to produce the observed 
\sit/\siw\ ratio, where $n_\gamma$ is the density of H-ionizing photons
($\lambda\!<\!912$\,\AA).
Second, we solve for the abundances of all the low-ion elements 
by comparing their observed columns with those predicted by the model. 

This method does not assume {\it a priori} that the heavy element abundances
are in their solar ratios; instead it solves for the abundance of each 
element separately, assuming the low-ions arise in the same gas phase as the 
\hi. See \citet{Ng12} for 
a discussion of effects occurring if clumpiness changes this assumption.
The models combine the UV background radiation calculated by \citet{HM12}
with a model of the escaping ionizing Milky Way radiation from \citet{Fo05a}, 
which was based on \citet{BM99, BM02}. 
The MW radiation field is non-isotropic, with the escape fraction highest 
normal to the Galactic disk, so $n_\gamma$ changes with latitude \citep{Fo05a}.

We ran {\it Cloudy} models for two of the four sightlines in this paper, \rbs\
(chosen to model the MS since it has the most UV metal 
lines detected) and \he\ (to model the AVCs near the South Galactic Pole).
For the \rbs\ sightline, we use a model with $l, b, R$=299.5\degr, 
--65.8\degr, 50\,kpc 
appropriate if the Stream is at the same distance as the Magellanic Clouds,
which gives log\,($n_\gamma$/cm$^{-3}$)=--5.61.
For the \he\ sightline, we adopt $l, b, R$=293.7\degr, --80.9\degr, 50\,kpc
giving log\,($n_\gamma$/cm$^{-3}$)=--5.34. 
We do not include the radiation escaping 
from the Magellanic Clouds 
or the radiation produced by the shock cascade or interface regions within 
the Stream. The inclusion of additional radiation fields would increase the 
photon density, so that for a given ionization parameter the clouds would 
become denser and smaller.

The results of our {\it Cloudy} models are given in two sets of figures.
First, Figure 9 shows the ionization corrections for six low ions calculated 
from the \emph{Cloudy} models, as a function of log\,$U$, for the \rbs\ and 
\he\ directions.
We also include a model appropriate to the \fn\ sightline, for use in the 
discussion and for comparison to Paper II.
The ionization corrections are defined as the difference between the 
intrinsic elemental abundance and the measured ion abundance, i.e.
IC(X$^{\rm i}$)=[X/H]--[X$^{\rm i}$/\hi]. 
Second, Figures 10a (for \rbs) and 10b (for \he) show the columns of the 
observed low ions as a function of log\,$U$, with the derived abundances 
of each element and the best-fit log\,$U$ annotated on the panel.
The results drawn from these two sets of figures are discussed in the next two 
sub-sections.

\begin{figure}[!ht]
\epsscale{1.25}
\plotone{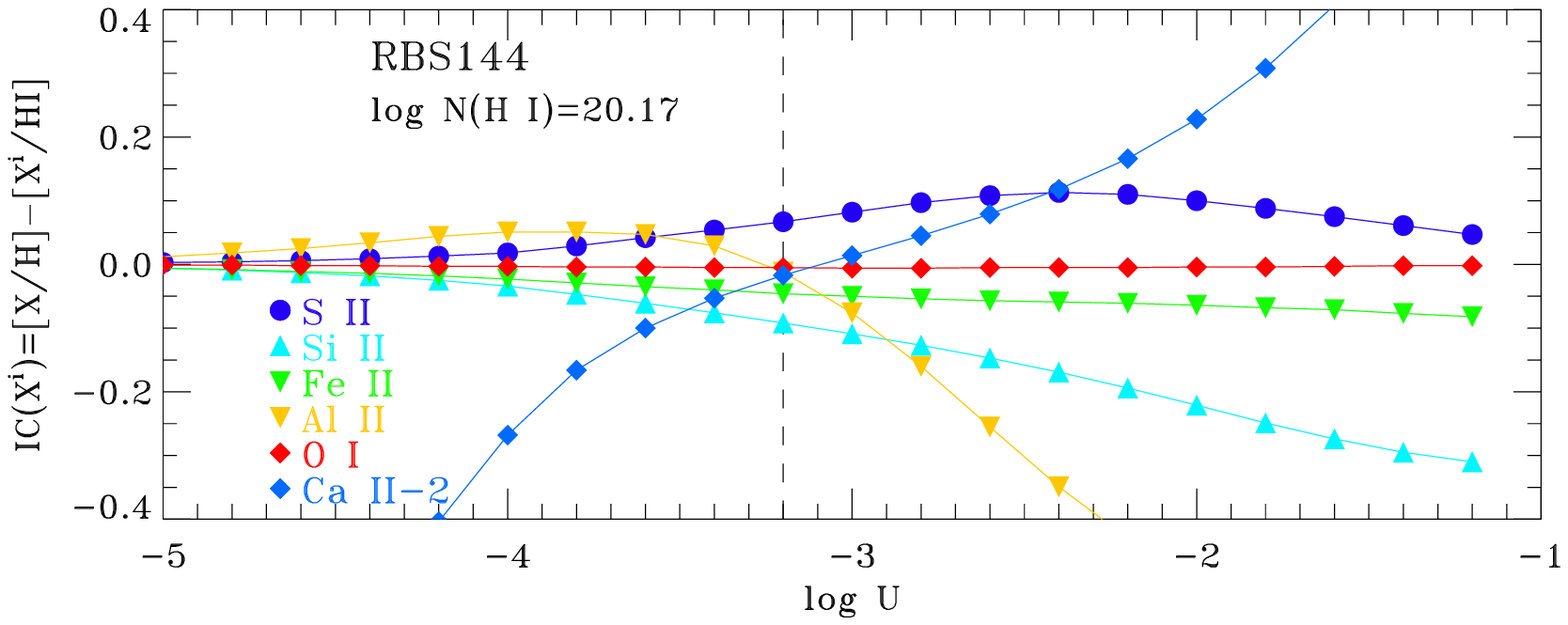}\plotone{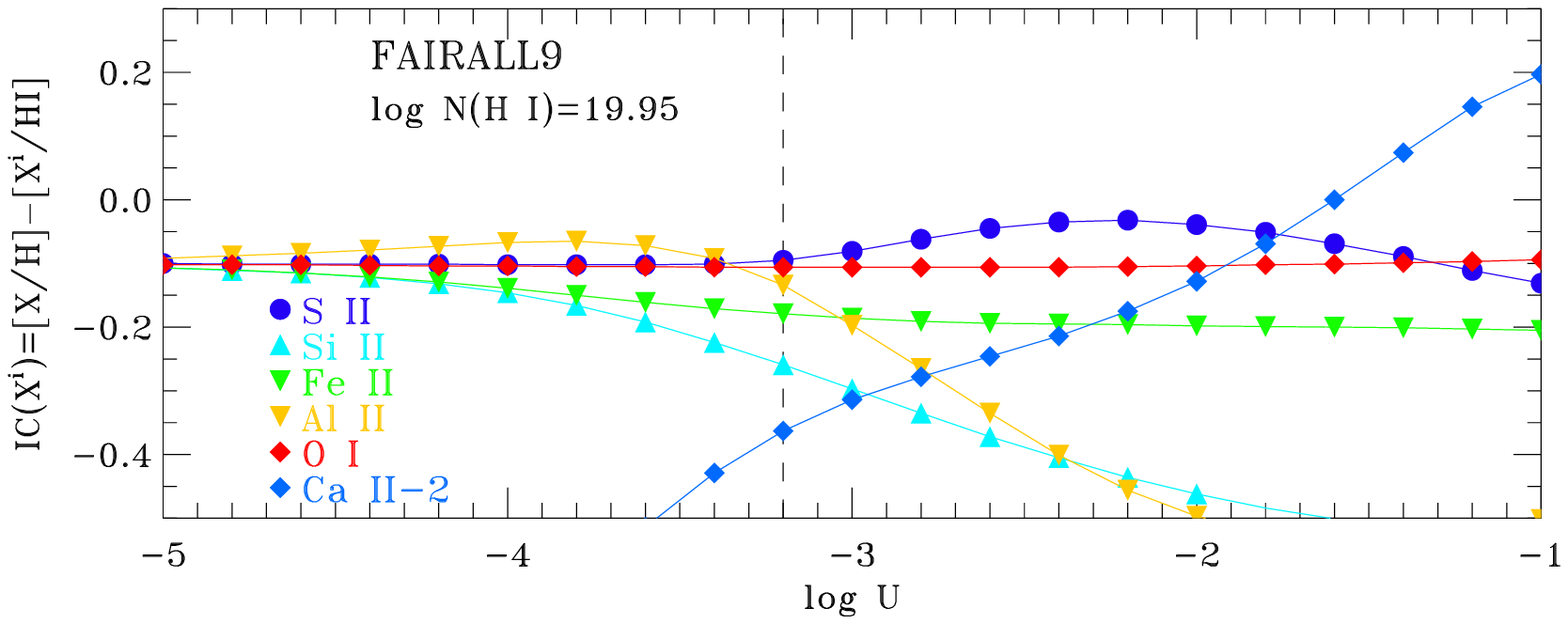}\plotone{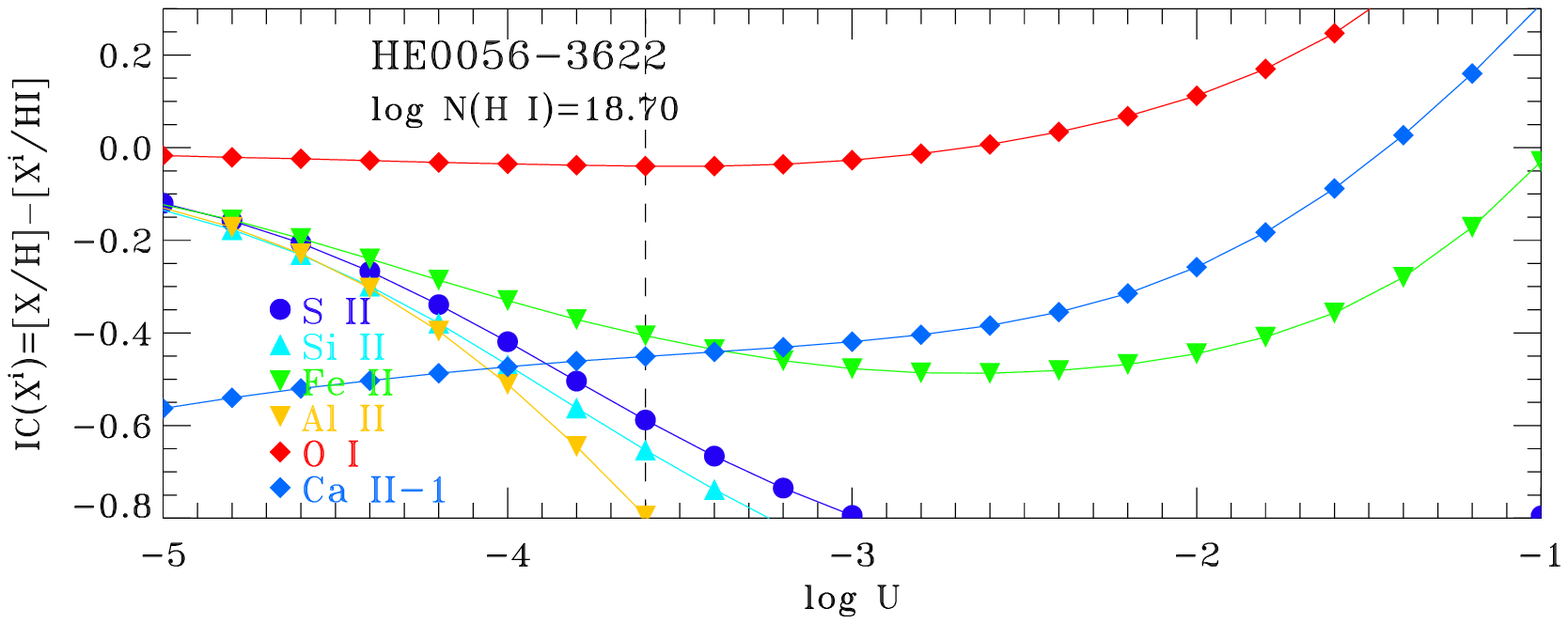}
\caption{Ionization corrections for six low ions (\oi, \sw, \siw, \alw, 
\few, and \caw) in our \emph{Cloudy} models of the MS toward 
RBS\,144 (top) and Fairall\,9 (center), 
and of the AVCs toward HE\,0056--3622 (bottom).
These corrections show the amount which must be added to the observed 
ion-to-\hi\ ratio to determine the intrinsic abundance. 
For RBS\,144 and Fairall\,9, the corrections for all ions shown are 
$\la$0.10\,dex at the best-fit value for log\,$U$ 
(derived by matching the \sit/\siw\ ratios), indicating that 
the sub-solar values derived for Si/S, Al/S, and Fe/S in the MS 
can be attributed to dust depletion, not ionization.
For HE\,0056--3622, the models support
the use of \oi/\hi\ as a robust abundance indicator even in the regime 
log\,$N$(\hi)$<$19, at least for log\,$U\la-2.0$.
Note that the \caw\ curves have been offset by --1.0 or --2.0\,dex on the 
y-axis for comparison to the other ions.}
\end{figure}

\begin{figure*}[!ht]
\epsscale{1.15}
\plottwo{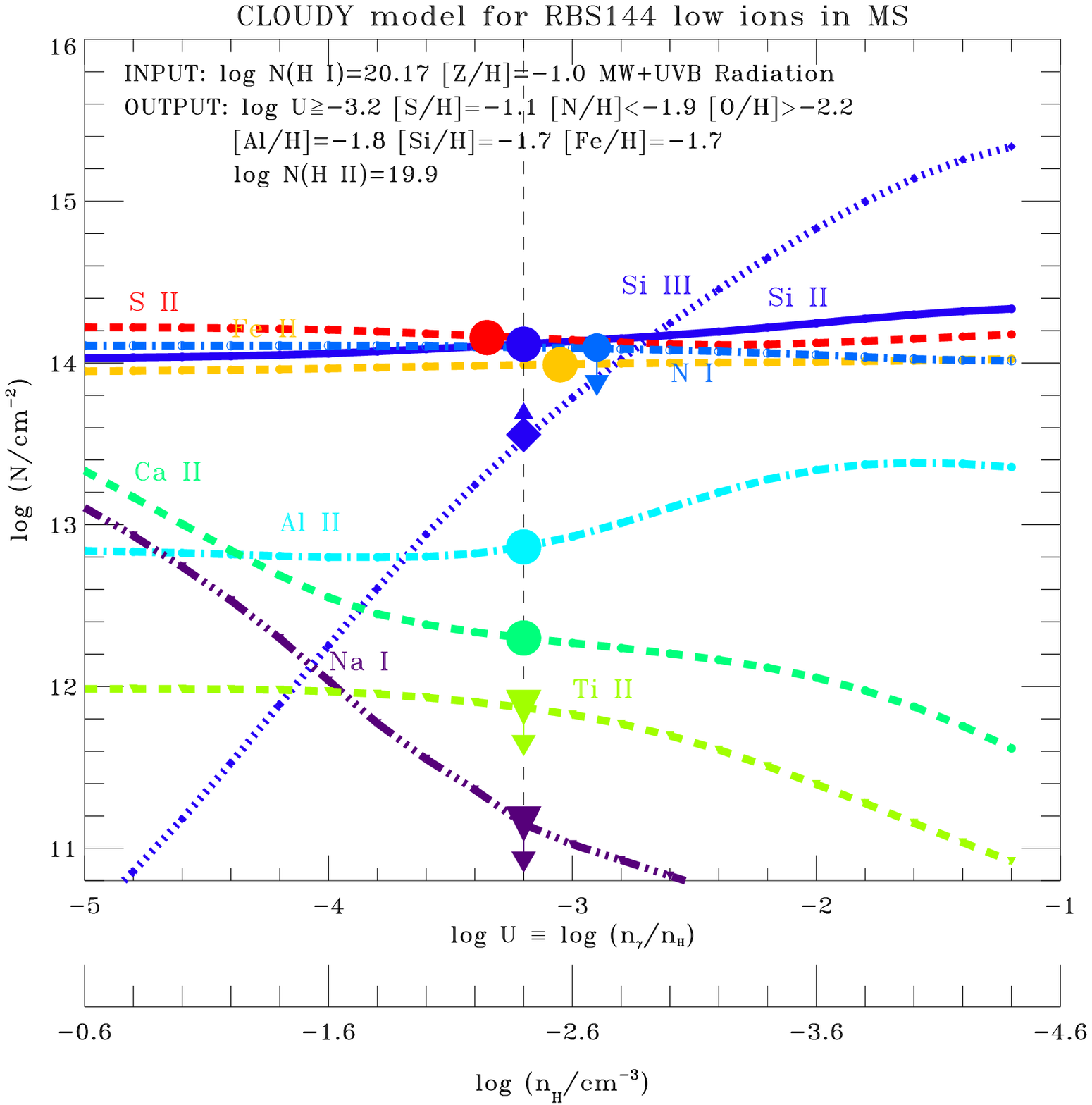}{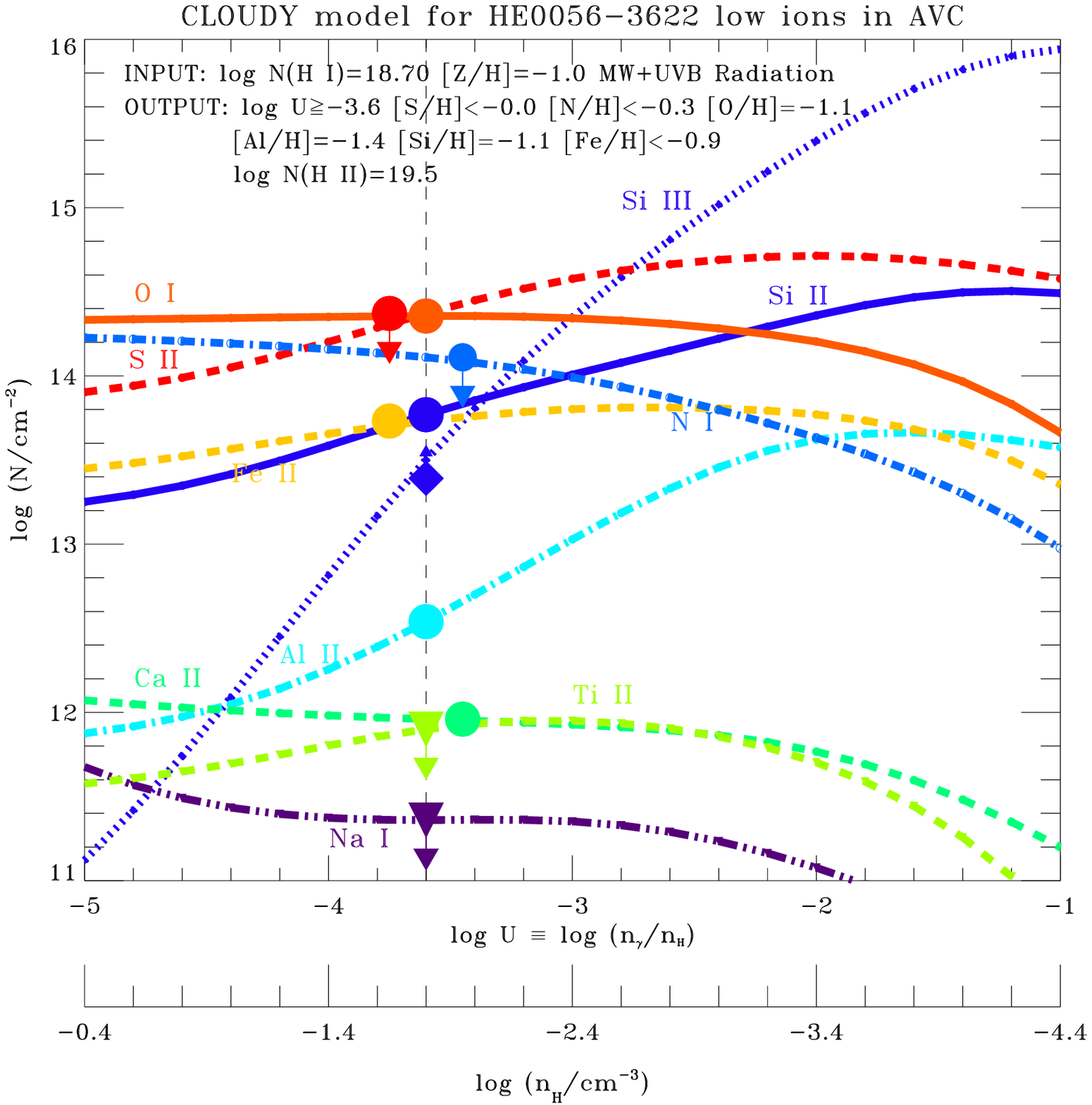}
\caption{Detailed \emph{Cloudy} photoionization models of the 
low-ionization species observed in the MS toward RBS\,144 (left), and 
in the AVCs toward HE\,0056--3622 (right). These models assume the 
low ions exist
in a single uniform-density phase exposed to the combined Milky Way 
plus extragalactic radiation field. The \sit/\siw\ ratio is used to derive 
the ionization parameter in the plasma (log\,$U\!\ga\!-3.2$ for the RBS\,144 
model and log\,$U\!\ga\!-3.6$ for the HE\,0056--3622 model).
The colored lines show the predicted column of each ion, scaled to match 
the observations at the best-fit log\,$U$; the amount by which each line
is scaled determines the gas-phase abundance of that element, as annotated 
on the plot. In turn, the comparison of the gas-phase abundances with the 
expected solar ratios indicates the depletion of each element onto dust grains.
Small offsets have been applied to the observations in the x-direction for
clarity. See \S4.3 for details.}
\end{figure*}

\subsubsection{Cloudy Results toward RBS\,144}
We measure a ratio log\,[$N$(\sit~1206)/$N$(\siw~1526)]$\ga$--0.56
over the full MS velocity interval 65--210\kms.
Reproducing this ratio with a \emph{Cloudy} model yields an ionization 
parameter log\,$U\!\ga\!-3.2$, corresponding 
to a gas density log\,($n_{\rm H}/{\rm cm}^{-3})\!\la\!-2.4$ (see Figure 10a).
These are limits since the MS component in \sit\ 1206 is saturated.
This model has an ionized gas column log\,$N$(\hw)=19.9.
If we repeat the measurement in the velocity interval 150 to 200\kms, where
\sit~1206 and \siw~1260 show a single, unsaturated component,
we derive log\,[$N$(\sit~1206)/$N$(\siw~1260)]=--0.13. 
Reproducing this ratio yields a best-fit log\,$U\ga-2.8$
and log\,($n_{\rm H}/{\rm cm}^{-3})\!\la\!-2.8$, so there is a $\approx$0.4\,dex
uncertainty in the best-fit density that results from saturation effects.

Regardless of the difficulty in determining the density to high precision,
the simulations confirm that the [\sw/\hi] ratio is a good 
indicator of the overall S abundance [S/H] in the MS: 
in our best-fit model along this sightline at log\,$U\!\ga\!-3.2$, 
the ionization correction is found to be IC(\sw)=0.07\,dex, 
and the correction is $<$0.1 dex \emph{over four orders of magnitude in density}
(Figure 9, top panel).
Physically, this is because the fraction of S in the form of S$^{+2}$ or higher
is similar to the fraction of H in the form of H$^+$
\citep[see][for related discussions]{Lu98, Ho06, HC12}.

The simulations confirm that Al, Si, and Fe are underabundant with respect 
to S in the MS, as we found in Figure 8 based on the ion abundances alone.
As for S, the ionization corrections for \alw, \siw, and \few\ 
derived by the \emph{Cloudy} model at log\,$U$=--3.2 are small:
IC(\alw)=--0.01\,dex, IC(\siw)=--0.09\,dex, and IC(\few)=--0.05\,dex, 
which is a consequence of the high \hi\ column in the Stream in this direction.
Applying these corrections we derive gas-phase abundances 
[Al/H]$_{\rm MS}$=--1.8, [Si/H]$_{\rm MS}$=--1.7, and [Fe/H]$_{\rm MS}$=--1.7, 
corresponding to moderate dust depletion levels
$\delta$(Al)$_{\rm MS}\!\approx\!-0.7$, $\delta$(Si)$_{\rm MS}\!\approx\!-0.6$, 
and $\delta$(Fe)$_{\rm MS}\!\approx\!-0.6$. 

In contrast, the derived Ca abundance from the \emph{Cloudy} model
is relatively high, [Ca/H]$_{\rm MS}$=--0.3, even though the \caw\
ion abundance is low, [\caw/\hi]=--2.23$\pm$0.10, because most of the 
Ca is predicted to be in the form of Ca$^{+2}$ (\ion{Ca}{3}). 
Equivalently, the \caw\ ionization correction in the model is large, 
IC(\caw)=+1.9. If correct, this would correspond to a negative Ca depletion
$\delta$(Ca)$_{\rm MS}$=--0.7 (i.e., a Ca enhancement relative to S), 
which would be puzzling for two reasons. First, Ca and S are both
$\alpha$-elements, so no nucleosynthetic difference in their abundances is
expected. Second, in the Galactic ISM, Ca readily depletes onto dust grains 
\citep{We96, SS96}, so \emph{sub-solar} Ca/S ratios are expected 
in the gas phase, 
not super-solar ratios. In another Stream direction (\object{NGC\,7469}), F10 
found [Ca/O]$\approx$0.0 after applying an ionization correction to the 
observed \caw/\hi\ ratio, so the issue is not unique to the \rbs\ direction.

The finding of no apparent Ca depletion in the Stream in both these directions
probably indicates that the single-phase assumption for the 
low ions, implicit within the \emph{Cloudy} models, breaks down, and that 
instead \caw\ preferentially traces regions of 
dense, cold neutral medium (CNM) which are not properly captured by 
our \emph{Cloudy} models, whereas \sw\ traces regions of warm neutral 
medium (WNM) and warm ionized medium (WIM). The observation that 
the \caw\ absorption in the MS toward \rbs\ occupies a far narrower velocity
interval than the other low ions supports this idea.
For \nao, the problem is even worse;
this ion has an ionization potential of only 5.1\,eV, so is destroyed easily
even in optically thick gas, and is therefore only able to survive in dense
clumps. We conclude that caution is necessary when using \emph{Cloudy} 
to model \nao\ and \caw\ in HVCs (and other ions whose ionization potential 
is less than 13.6\,eV), since they likely do not coexist in the same
gas phase as the \hi\ and the other low ions 
\citep[see Paper II and][for more discussion on this point]{Ri05,Ri11}.

\subsubsection{Cloudy Results toward HE\,0056--3622}
Integrating over the full AVC velocity interval 80--200\kms, we measure
a ratio log\,[$N$(\sit\ 1206)/$N$(\siw\ 1526)]=--0.38. 
Matching this ratio gives log\,$U\!\ga\!-3.6$, corresponding to a best-fit 
density log\,($n_{\rm H}/{\rm cm}^{-3})\!\la\!-1.8$ (see Figure 10b). 
At this log\,$U$, the model has an ionized gas column log\,$N$(\hw)=19.5,
and an ionization fraction \hw/(\hi+\hw)=86\%. 
If we adopt the ratio log\,[$N$(\sit)/$N$(\siw)]=--0.90 measured
over the smaller interval 150--200\kms, 
where both \sit\ 1206 and \siw\ 1526 are 
unsaturated, we derive log\,$U$=--4.0 and log\,($n_{\rm H}/{\rm cm}^{-3})$=--1.4,
indicating that the error in the gas density arising 
because of possible saturation of \sit\ 1206 is $\approx$0.4\,dex. 

The model confirms that [\oi/\hi]=[O/H] for all ionization parameters 
log\,$U\!\la\!-2.5$, even at the low neutral gas column 
log\,$N$(\hi)$_{\rm AVC}$=18.70 present in this sightline \citep[see][]{Vi95}. 
The ionization corrections for \alw, \siw, and \few\ 
at the best-fit log\,$U$=--3.6 are more significant in this sightline 
than toward \rbs\ due to the lower \hi\ column (Figure 9, bottom panel):
IC(\alw)=--0.8\,dex, IC(\siw)=--0.6\,dex, and IC(\few)=--0.4\,dex. 
Using these corrections we derive gas-phase abundances 
[Al/H]$_{\rm AVC}$=--1.4, [Si/H]$_{\rm AVC}$=--1.1, and [Fe/H]$_{\rm AVC}\!<\!-0.9$,
corresponding to (small) depletions relative to oxygen of 
$\delta$(Al)$_{\rm AVC}\!\approx\!-0.2$, 
$\delta$(Si)$_{\rm AVC}\!\approx\!0.0$, and $\delta$(Fe)$_{\rm AVC}\!<\!+0.1$. 
There is therefore no evidence for dust depletion in the AVC toward \he. 

\section{Discussion}
The one-tenth-solar metallicity in the Stream measured from 
[\sw/\hi] toward \rbs\ and from [\oi/\hi] toward \ngc\
matches the value measured from [\oi/\hi] toward \object{NGC\,7469} by F10. 
These measurements are shown in Figure 11, where we plot MS metallicity
against angular distance from the center of the SMC
for four MS directions observed as part of our \hst/COS program,
and one (\object{NGC\,7469}) observed with \hst/STIS.
{\it There are therefore three independent measurements of 
one-tenth-solar metallicity along the main body of the MS, and a fourth
(\phl) with an upper limit consistent with one-tenth solar.}
However, these measurements differ from the much higher value 
[\sw/\hi]$_{\rm MS}$=--0.30$\pm$0.04 (0.5 solar)  
measured toward \fn\ \citep[Paper II;][]{Gi00}, lying
only 8.3\degr\ away on the sky from \rbs, in a direction with a fairly 
similar MS \hi\ column density (19.95 for \fn\ vs 20.17 for \rbs). 
This corresponds to an abundance variation of a factor 
of five over a small scale.
However, the Stream's LSR velocity centroid changes by $\approx$100\kms\
between these two directions, from 92\kms\ toward \rbs\ to 194\kms\ toward 
\fn, indicating that the Stream in the two directions also has very
different \emph{kinematic} properties. 

\begin{figure*}[!ht]
\epsscale{1.25}
\plotone{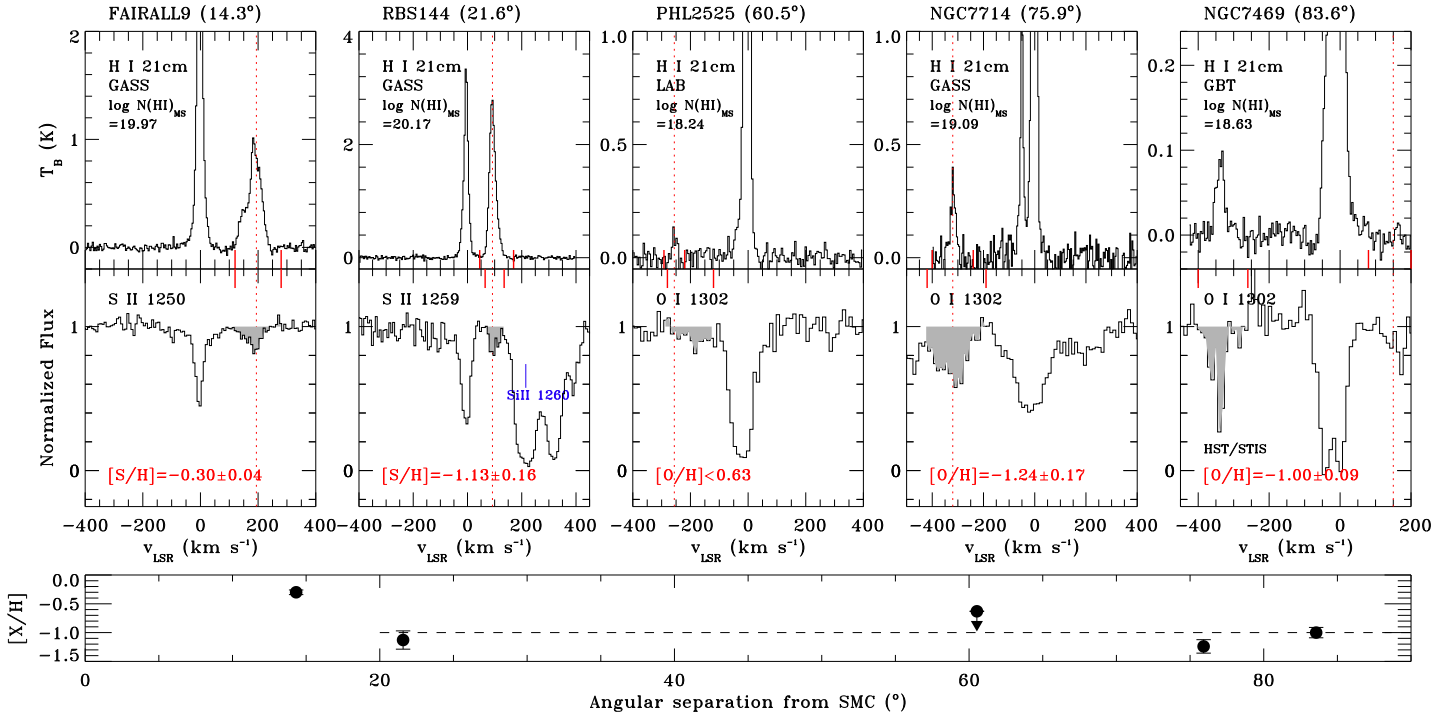}
\caption{Summary of the metal abundance pattern along the body of the Stream.
The upper panels show the GASS, LAB, or Green Bank Telescope (GBT)
21\,cm profiles, giving the \hi\ columns 
in the MS. The middle panels show the profiles of \oi\ 1302, 
\sw\ 1259, or \sw\ 1253, giving the metal-line columns. 
The numbers in parentheses in the title of each column show the angular 
separation of each direction from the center of the SMC, so the plot is 
presented in order of increasing separation from the SMC. 
The lower panel shows the MS abundance [X/H] (where X is S or O) versus
separation from the SMC. The abundance is effectively constant at $\approx$0.1 
solar along the main body of the Stream (shown with the dashed line), 
until it jumps sharply in the Fairall\,9 direction (Paper II). 
The NGC\,7469 data are from F10.}
\end{figure*}

Indeed, if we place our targets on the N08 map of the two bifurcated filaments
of the Stream, we find that \fn\ lies behind the LMC filament,
whereas \rbs\ lies behind the {\it second} filament, matching its position both
spatially and kinematically (see Figure 1).
Thus our 0.1\,solar metallicity measurement in the \rbs\ direction indicates an 
SMC origin for this second filament (for the reasons given in the next 
paragraph). The 0.5\,solar metallicity measurement in the \fn\ direction is 
consistent with the LMC origin for that filament reported by N08, 
although the low N/$\alpha$ ratio in that sightline
and its proximity to both Magellanic Clouds suggests a complex
metal enrichment history (Paper II).
This is particularly true in light of recent results that both Magellanic Clouds
experienced a burst of star formation $\approx$2\,Gyr ago \citep{We13}.

Our finding that the main body of the Stream has a metal abundance close 
to 0.1 solar in three directions strongly supports the tidal origin 
scenario in which most of the Stream was stripped from the 
SMC $\approx$2\,Gyr ago. This is because 
(1) the SMC's age-metallicity relation \citep{PT98, HZ04} indicates that its 
average abundance 2\,Gyr ago was $\approx$0.1--0.15 solar, and
(2) the tidal age of the Stream in many numerical studies is 
$\approx$1.5--2.5\,Gyr \citep{Li95, Ga94, GN96, Co06, Be10, Be12, DB11a, DB12}. 
That is, the Stream's tidal age and its chemical age are both consistent with 
an SMC origin. The gas-to-dust ratio measured in the MS toward \rbs\ also 
agrees with this scenario, lying within a factor of two of the current-day 
gas-to-dust ratio in the SMC.
In contrast, our low MS abundance measurements in all directions except 
\fn\ indicate the main body of the Stream is chemically inconsistent with an 
LMC origin, since it has been $\approx$5--10\,Gyr since the average LMC 
abundance was as low as 0.1 solar \citep{PT98, HZ09, BT12}, and the Stream is 
not that old in any realistic model, except for the regions closest to the 
Magellanic Clouds which may have been released or polluted more recently.
We caution that any metallicity gradients present in the LMC or SMC complicate
this interpretation, since lower-metallicity gas at the outskirts of either 
galaxy will be easier to strip (either by ram pressure or by tidal forces) 
than gas in the inner regions.

It is worth considering whether some of the difference between the 
Stream's metallicity in the \rbs\ and \fn\ directions
could be attributed to small-scale (sub-resolution) structure in \hi,
which is known to be present in the Stream 
\citep{Pu03a, St02, St08, WK08, Ni10}.
\citet{Wa11} have shown that for radio telescopes with beam-sizes of order 
10\arcmin, variations of up to 20--25\% in $N$(\hi) determinations can occur. 
In F10, we compared $N$(\hi)$_{\rm MS}$ determinations toward \object{NGC\,7469}
from four radio telescopes of beam sizes ranging from 9.1\arcmin\ to 35\arcmin,
and found the values varied by 0.15\,dex (40\%).
While important, these uncertainties are far less than the factor of five 
difference in MS metallicity we observe between the \rbs\ and \fn\ directions.
Furthermore, for \rbs\ the pencil-beam \hi\ column derived from \lya\
agrees within 2$\sigma$ with the \hi\ column derived from GASS,
and so the metallicities derived in this direction do not have a 
large systematic beamsize error. We conclude that the higher-velocity 
filament of the MS seen toward \fn\ has a genuinely higher metallicity than the 
lower-velocity filament seen toward \rbs, despite their proximity on the sky. 

Finally, we note that our 0.1\,solar metallicity measurements in most of the MS
are consistent with the abundances measured in the Magellanic Bridge (MB) of 
gas connecting the LMC and SMC. \citet{Le08} measured 
[O/H]$_{\rm MB}$ =--0.96$^{+0.13}_{-0.11}$ toward O-star \object{DI~1388},
and [O/H]$_{\rm MB}$=--1.36 toward O-star \object{DGIK~975}, although these stars 
appear to lie in front of the principal \hi-emitting phase of the Bridge, 
so only sample the foreground part of it. 
Our MS metallicity measurements are also consistent with
the results of \citet{Mi09}, who reported --1.0$<$[Z/H]$_{\rm MB}$$<$--0.5 toward 
QSO \object{PKS\,0312--770}, the sightline to which samples the full radial
extent of the Bridge.
However, we note that in recent simulations \citep{Be12, DB12}
the Bridge and the Stream are created at different points in time,
so their similar present-day metallicities may be partly coincidental.

\section{Summary}
In this first paper of a series presenting combined \hst/COS and 
VLT/UVES spectroscopy of the Magellanic Stream, we have investigated the 
Stream's chemical abundances using spectra of four AGN:
\rbs, \phl, \ngc, and \he. 
These sightlines all lie behind the main body of the Stream,
and sample a wide range of \hi\ column density, from 
log\,$N$(\hi)$_{\rm MS}$=20.17 toward \rbs\ down to 18.24 toward \phl.
Three of these targets (\rbs, \phl, and \ngc) have UV spectra that 
allow MS abundance measurements.
The fourth (\he) cannot be used for this purpose since
the MS velocity centroid in this direction (--10\kms) overlaps
with foreground Galactic absorption; however, we have measured the 
anomalous velocity cloud (AVC) absorption centered at 150\kms\ in 
that sightline. A wide range of metal-line species is detected in the Stream 
including \oi, \cw, \cf, \siw, \sit, \sif, \sw, \alw, \few, and \caw.
Our main results are as follows.

\begin{enumerate}

\item
Toward \rbs, we measure a MS sulfur abundance 
[S/H]$_{\rm MS}$=[\sw/\hi]$_{\rm MS}$=--1.13$\pm$0.16. 
Toward \ngc, we measure a similar value
[O/H]$_{\rm MS}$=[\oi/\hi]$_{\rm MS}$=--1.24$\pm$0.20. 
Toward \phl, we measure an upper limit
[O/H]$_{\rm MS}\!<\!-0.63$ (3$\sigma$), based on a non-detection in \oi\ 1302. 
When combining these new measurements
with the published metallicity toward \object{NGC\,7469} (F10),
there are now three good measurements 
along the main body of the MS consistent with $\approx$0.1 solar metallicity.
This matches the SMC's abundance
$\approx$2\,Gyr ago, given its age-metallicity relation \citep{PT98,HZ04}.
Tidal origin models for the Stream {\it independently} age it 
at $\approx$1.5--2.5\,Gyr
\citep[e.g.][]{Ga94, Co06, DB11a, DB12, Be12}.
Furthermore, the gas-to-dust ratio we measure in the Stream toward
\rbs\ is within a factor of two of the average value found in the 
diffuse gas of the SMC. 
{\it Thus we conclude based on chemical evidence
that most of the Stream originated in the SMC about 2\,Gyr ago.} 
We note that the \rbs\ sightline passes through the second (non-LMC) filament 
identified in \hi\ 21\,cm emission by N08, and thus our metallicity measurements
resolve the nature of this filament's origin (in the SMC). 

\item
The Stream's [S/H] abundance we measure toward \rbs\ differs by 
$\approx$0.7\,dex (a factor of five)
from the value [S/H]$_{\rm MS}$=--0.30$\pm$0.04 measured toward \fn\ 
(only 8.3\degr\ away) in Paper II of this series. 
\fn\ lies behind the LMC filament of the Stream (N08). 
Our \emph{Cloudy} simulations indicate that ionization corrections are not 
the reason for the different abundances:
we find that the ionization correction
IC(\sw)=[S/H]--[\sw/\hi]$\approx$0.07\,dex at the best-fit density 
toward \rbs, and $|$IC(\sw)$|\!<\!0.01$\,dex toward \fn, indicating
the difference in the abundances is real.
Furthermore, the Stream exhibits very different gas-to-dust 
ratios and velocity centroids in the two directions. 
{\it This shows that the bifurcation of the Stream
is seen not only in its spatial extent and its kinematics,
but also in its chemical enrichment.}

\item
We find evidence for dust depletion in the Stream in the form of
sub-solar Si/S, Al/S, and Fe/S ratios toward \rbs,
where S is used as an undepleted reference element.
Toward \rbs, we measure depletion levels of 
$\delta$(Al)$_{\rm MS}\!\approx\!-0.7$, 
$\delta$(Si)$_{\rm MS}\!\approx\!-0.6$, and $\delta$(Fe)$_{\rm MS}\!\approx\!-0.6$.
These depletions are similar to those found in the Leading Arm by \citep{Lu98},
and indicate that dust grains survive the process that ejected the Stream
from the Magellanic Clouds.

\item
Toward \he, we measure an oxygen abundance
[O/H]$_{\rm AVC}$=[\oi/\hi]$_{\rm AVC}$=--1.03$\pm$0.18 
in the cloud centered at $v_{\rm LSR}$=150\kms, 
which belongs to a population of anomalous velocity clouds (AVCs) found near
the South Galactic Pole. The similarity between this metallicity and the MS 
metallicity suggests the AVCs are associated with the Stream rather than 
tracing foreground Galactic material. 
However, we cannot rule out the possibility that they are associated with more
distant material, e.g. in the Sculptor Group, which lies in this direction
at similar velocities \citep{Ma75, Pu03a}. 

\end{enumerate}

A detailed study of the Stream's chemical abundances in the \fn\ direction
is presented in Paper II. In Paper III we will address
the Stream's ionization level and fate, and
the more general question of the role of tidal streams in fueling
$L_\ast$ galaxies.\\

{\it Acknowledgments}\\
We gratefully thank Justin Ely for providing a night-only reduction of the 
COS data, and we acknowledge valuable conversations with Julia Roman-Duval,
David Nidever, Gurtina Besla, Karl Gordon, Mary Putman, and Kat Barger.
We are grateful to the referee for a useful and insightful report. 
Support for program \#12604 was provided by NASA through a grant from 
the Space Telescope Science Institute, which is operated by the Association
of Universities for Research in Astronomy, Inc.,
under NASA contract NAS 5-26555.

\end{document}